\documentclass[12pt, a4paper, table, xcdraw]{article}
\pdfoutput=1
\usepackage{geometry}
\usepackage{amsmath, amssymb, bbm, bm}
\usepackage{cite}
\usepackage{graphicx}
\usepackage{caption, subcaption}
\usepackage[T1]{fontenc}
\usepackage{footnote}
\usepackage{pdfpages}
\usepackage{hhline}
\usepackage{multirow, multicol}
\usepackage{colortbl} 
\usepackage{xcolor}   
\usepackage{enumitem}
\usepackage{relsize}
\usepackage[toc,page]{appendix}
\usepackage[english]{babel}
\allowdisplaybreaks
\usepackage{xcolor}
\usepackage{float}
\usepackage{slashed}
\usepackage{cases}
\usepackage{empheq}
\usepackage{hyperref}
\hypersetup{
    colorlinks=true,
    linkcolor=MyGreen,
    citecolor=SBlue,
    urlcolor=MyDarkBlue,
    bookmarksnumbered=true,
    bookmarksopen
}

\definecolor{MyDarkBlue}{rgb}{0.1, 0.3, 0.8}
\definecolor{SBlue}{rgb}{0.2, 0.4, 0.4}
\definecolor{MyLightBlue}{rgb}{0.22, 0.51, 0.99}
\definecolor{MyGreen}{rgb}{0.0, 0.5, 0.3}
\definecolor{BrickRed}{rgb}{0.8, 0.25, 0.33}
\definecolor{lightblue}{rgb}{0.8, 0.9, 1.0}
\definecolor{lightgreen}{rgb}{0.9, 1.0, 0.9}
\definecolor{lightyellow}{rgb}{1.0, 1.0, 0.8}
\definecolor{lightgray}{rgb}{0.9, 0.9, 0.9}

\newcommand\identity{1\kern-0.25em\text{l}}

\renewcommand\thesection{\arabic{section}.}
\usepackage{etoolbox}
\AtBeginEnvironment{appendices}{%
    \addtocontents{toc}{\protect\renewcommand{\protect\addcontentsline}[3]{}}
}

\begin{document}

\vspace*{0.3in}
\begin{flushright}
\end{flushright}
\begin{center}
{\large \bf Radiative Origin of Fermion Mass  Hierarchy \\\vspace{0.01in} in Left-Right Symmetric Theory
}
\end{center}
\renewcommand{\thefootnote}{\fnsymbol{footnote}}
\begin{center}
{
{}~\textbf{Sudip Jana$^1$,}\footnote{ E-mail: \textcolor{MyDarkBlue}{sudip.jana@mpi-hd.mpg.de}}
{}~\textbf{Sophie Klett$^1$,}\footnote{ E-mail: \textcolor{MyDarkBlue}{sophie.klett@mpi-hd.mpg.de}}
{}~\textbf{Manfred Lindner$^1$,}\footnote{ E-mail: \textcolor{MyDarkBlue}{lindner@mpi-hd.mpg.de}}
{}~\textbf{Rabindra N. Mohapatra$^2$}\footnote{ E-mail: \textcolor{MyDarkBlue}{rmohapat@umd.edu}}
}
\vspace{0.5cm}
{
\\\em $^1$ Max-Planck-Institut f{\"u}r Kernphysik, Saupfercheckweg 1, 69117 Heidelberg, Germany
\\ $^2$ Maryland Center for Fundamental Physics and Department of Physics, University of Maryland, College Park, Maryland 20742, USA
}
\end{center}
\renewcommand{\thefootnote}{\arabic{footnote}}
\setcounter{footnote}{0}
\thispagestyle{empty}

\begin{abstract} 
Despite the remarkable success of the Standard Model, the hierarchy and patterns of fermion masses and mixings remain a profound mystery. To address this, we propose a model employing the rank mechanism, where the originally massless quarks and leptons sequentially get masses. The third generation masses originate from the seesaw mechanism at the tree-level, while those of the second and first generations emerge from one-loop and two-loop radiative corrections, respectively, with a progressive increase in the rank of the mass matrix. This approach does not require new discrete or global symmetries.  Unlike other theories of this type that require the introduction of additional scalars, we employ the double seesaw mechanism within a left-right symmetric framework, which allows us to realize this scenario solely through gauge interactions. 
\end{abstract}
\newpage
\setcounter{footnote}{0}
{
  \hypersetup{linkcolor=black}
  \tableofcontents
}
\newpage

\section{Introduction}\label{SEC-00}
The Standard Model (SM) of particle physics is very successful, but many important questions are left unanswered. A primary one among them is the nature of fermion flavor i.e. why the quark and charged lepton masses are hierarchical and why are the quark mixings also similarly nearest neighbor type. The neutrinos exhibit a much weaker hierarchy, with mixings almost anarchical. A great deal of effort has gone into understanding these features within the framework of gauge theories, with most approaches using extra discrete or global family symmetries~\cite{Froggatt:1978nt, Arkani-Hamed:1999ylh, Babu:2009fd, King:2013eh}, that often lead to new particles called flavons and suitable symmetry breaking sequences. The question then comes up as to how to understand the origin of these symmetries and of their breaking. 

Are new family symmetries really necessary? Where are the flavons, if they exist?  Clearly, an explanation of the mass hierarchies without extraneous symmetries would bypass such questions and provide a far cleaner route. One such approach was proposed in a series of papers in the late 1980s and has been followed up subsequently~\cite{ Balakrishna:1987qd, Balakrishna:1988ks, Balakrishna:1988bn, Babu:1989tv, Babu:1988fn, Rattazzi:1990wu, Dobrescu:2008sz, Weinberg:2020zba, Mohanta:2022seo, Mohanta:2023soi,  Graham:2009gr, Jana:2021tlx,Bonilla:2023wok ,Mohanta:2024wcr}. 
The strategy employed in this approach is the following. It is well known that in the limit of zero Yukawa couplings, the standard model has a $U(3)_Q\times U(3)_\ell \times U(3)_{u_R}\times U(3)_{d_R}\times U(3)_{e_R}$ symmetry. It is generally believed that observed fermion masss hierarchy is related to the breaking of this symmetry. We use a framework where we exploit this feature and include heavy vector-like fermions in such a way that the above symmetry only holds for the usual quarks and leptons and not for the full Lagrangian. The symmetry is then partially broken at the tree-level by certain Yukawa couplings involving SM fermions with heavy vector like fermions. This gives mass only to the third generation generation fermions at the tree-level with the heavy vector-like fermion dynamics breaking the rest of the symmetry in higher orders, which then generates masses for the second and first generation fermions. We call this the rank mechanism and no flavons are needed.


To implement this strategy, it is convenient to use the so-called universal seesaw mechanism \cite{Berezhiani:1983hm, Chang:1986bp, Davidson:1987mh, Rajpoot:1987fca, Babu:1988mw ,Babu:1989rb, Mohapatra:2014qva, Patra:2017gak, Chen:2022wvk, Dcruz:2022rjg, Morozumi:2024mit}, within the framework of standard model (SM) or the left-right symmetric models (LRSM). Various scenarios implementing this strategy, that have been discussed in the literature extend the SM or the LRSM models with new vectorlike fermions with interactions involving scalars, with new Yukawa type interactions. Although they reproduce the mass hierarchy, they generally involve many new parameters. It is evident that gauge interactions would typically provide a more streamlined and economical alternative in terms of overall calculability and productivity. A notable instance of this is the framework suggested in \cite{Weinberg:2020zba}, where the highly predictive nature of the proposed implementation was ultimately dismissed due to inconsistencies with the experimentally observed masses and mixing parameters of charged fermions. Our approach in this paper differs from this as well as other theories which generally require additional scalars. We show that we can implement the rank mechanism by introducing only extra vector-like fermions and using a double seesaw mechanism without the need for any extra scalars to explain the fermion mass hierarchy. As a result, in the unitary gauge, the interactions contributing to loops involve only gauge fields and three neutral scalars, including the 125 GeV SM Higgs field and only their interactions. They seem to be sufficient to explain the observed fermion mass hierarchy. Unlike traditional left-right symmetric theories, where the masses of all charged fermions result from either a bidoublet Higgs field or a generalized seesaw mechanism \cite{Davidson:1987mh} at the tree-level, our proposal uses neither of them and relies exclusively on partial seesaw supplemented by dynamics of gauge interactions. New vector-like top and bottom-like quarks ($T$, $B$) and charged leptons ($E$) generate the masses of third-generation fermions at tree-level.  The gauge interactions are, then, used to generate the masses of first and second-generation fermions at two-loop and one-loop levels, respectively, with loop factors providing the mass hierarchy. This implements the rank mechanism. The pattern of quark mixings, however, is not explained, although we give a fit to quark mixings using parameters of the theory. Even though our model is completely quark lepton symmetric, the neutrino sector in these models always needs a separate treatment, due to the huge mass gap between charged fermions and neutrinos. Neutrinos being neutral fermions allow more kinds of couplings and in this paper, we will use high scale lepton number violating terms to understand neutrino masses. 

This paper is structured as follows: the next section introduces the rank mechanism. Following this, we present the left-right symmetric model incorporating new fermions. The following section delves into the generation of quark masses at tree-level, one-loop, and two-loop levels, followed by an analysis of lepton mass generation. In the subsequent section, we provide numerical results. Finally, we conclude.

\section{Rank mechanism for SM fermion mass hierarchy}\label{SEC-01}
Left-right symmetric models (LRSM)~\cite{Pati:1974yy, Mohapatra:1974hk, Mohapatra:1974gc, Senjanovic:1975rk},
based on the gauge group $\mathcal{G}_{\mathrm{LRSM}} = SU(3)_C \times SU(2)_R \times SU(2)_L \times U(1)_{X} $\footnote{In \cite{Babu:2022ikf} it is discussed that $X$ could be identified with $B-L$, however this would mean that $B-L$ appears to be broken when SM fermions mix with the vector-like fermions that carry a different $B-L$ charge. },  constitute a natural framework to realize the idea of mass matrices with a seesaw-like texture for all generations and have been extensively studied in previous works \cite{Babu:1989rb, Babu:1988mw, Babu:2018vrl}. In these models, left-handed (right-handed) SM fermions belong to a $SU(2)_L$ $\left(SU(2)_R \right)$ doublet representation and the theory contains two scalars $\chi_L$ and $\chi_R $, doublet under $SU(2)_L$ and $SU(2)_R$, respectively. Without a scalar bi-doublet, the usual Yukawa couplings between left -and right-handed fermion doublets are forbidden. Extending the particle content by $m$ generations of massive vector-like fermions that are singlet under $SU(2)_L \times SU(2)_R$ then naturally gives rise to seesaw-textured mass matrices 
\begin{equation}
\label{Eq.: Universal Seesaw}
    \mathcal{M} = \left(
    \begin{array}{cc}
        \mathbf{0}& v_L |\alpha\rangle  \\
        v_R \langle \alpha| &M_T
    \end{array}
    \right) ~,
\end{equation}    
where $|\alpha\rangle$ includes the Yukawa couplings, $v_L$, $v_R$ denote the vacuum expectation values (VEVs) of the scalar fields $\chi_L$, $\chi_R$, and $M_T$ is a vector-like mass matrix of dimension $m\times m$. Since $\mathcal{M}$ has rank $2m$, the number of vector-like fermion generations provides control over the number of massive SM generations at the tree-level. 

We seek to find a model that realizes the fermion mass hierarchy due to successive loop suppression by using only the gauge and doublet scalar interactions and without adding any extra scalars beyond those that are necessary to implement gauge symmetry breaking. The inclusion of these loop effects should steadily increase the matrix rank until all fermions obtain a mass. However, with generation independent fermion charges and flavor diagonal gauge couplings, typical one-loop corrections to the zero-entry of $\mathcal{M}$ are of the form $\delta M_{11} \sim |\alpha\rangle \langle \alpha|$ and hence cannot increase the rank \cite{Jana:2021tlx, Mohanta:2022seo}. To solve this issue, we consider a generic $(3+m+n) \times (3+m+n) $ mass matrix of the form
\begin{equation}
\label{Eq.: Double Seesaw}
    \mathcal{M} = \left(
    \begin{array}{ccc}
        \mathbf{0}& v_L |\alpha\rangle&\mathbf{0}  \\
        v_R \langle \alpha| &M_T &  |\beta\rangle  \\
       \mathbf{0}&  \langle \beta| & \mathbf{0}
    \end{array}
    \right) ~,
\end{equation}  
with some $m \times n$ vector $|\beta\rangle$. The tree-level rank of this matrix is $2m$ if we include the left and right helicity fermions separately. Though, within this extended setting, a one-loop correction to the upper left zero-entry given by $\delta M_{11} \sim |\alpha\rangle \langle \alpha|$ increases the matrix rank by one, given $m < 3+n $.\footnote{If $m \geq 3+n$ the matrix would be already a full rank matrix and enhancing the number of non-zero eigenvalues is not possible anymore.} Note that this feature was already explored in the context of radiative inverse seesaw models for neutrinos \cite{Dev:2012sg}. If, at the two-loop level, any kind of independent new coupling is introduced, the matrix rank increases again. The aim of the following work is to give a concrete example of this idea within the framework of left-right symmetric models with $m=2$ and $n=1$. While the third generation of SM fermions obtains a mass from a tree-level seesaw with vector-like new fermions, the second generation becomes massive at the one-loop level by radiative neutral gauge boson corrections and first-generation masses arise only at the two-loop level by $W_L-W_R$ exchange (which mix at one-loop) and encompass new independent couplings from the $SU(2)$ isospin partners.
Note that the mass hierarchy is generated only by gauge interactions, and no new scalars are introduced.
\begin{figure}[htb!]
	\centering
	\includegraphics[scale=0.22]{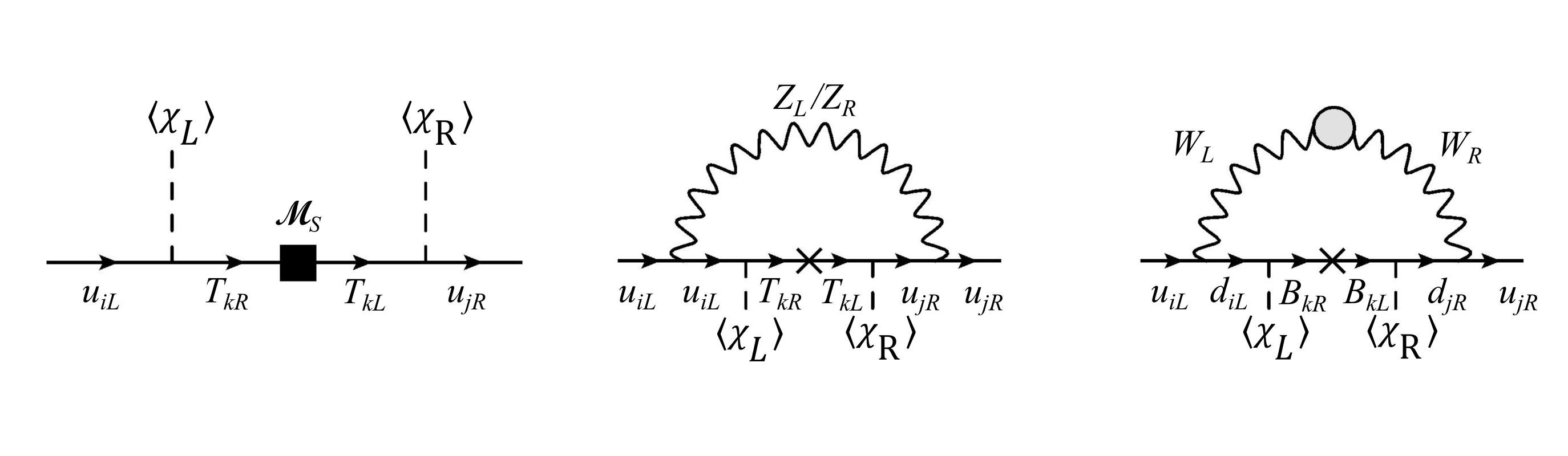}
	\caption{Schematic of mass generation for quarks and charged leptons with third generation masses at tree-level (left),   second generation masses at one-loop (middle), and first-generation masses generated at the two-loop level (right).}
	\label{Fig.: Schematic mechanism}
\end{figure}
\section{Left-right symmetric model with new fermions}\label{SEC-02}
We start by considering the type of LRSM where fermion masses originate from a universal seesaw mechanism  \cite{Berezhiani:1983hm, Chang:1986bp, Davidson:1987mh, Rajpoot:1987fca, Babu:1988mw, Babu:1989rb}. In this class of models, the SM fermions and three right-handed neutrinos are assigned to the following representations under $\mathcal{G}_\mathrm{LRSM}$ ($j=1,2,3$ is a family index)
\begin{equation*}
    \begin{split}
        Q_{jL} = \left(
		\begin{array}{c}
			u_j\\
			d_j\\
		\end{array} \right)_L \sim (3,2,1,1/3)~, ~ 
        Q_{jR} = \left(
		\begin{array}{c}
			u_j\\
			d_j\\
		\end{array} \right)_R \sim (3,1,2,1/3)~, ~  \\
       \Psi_{jL} = \left(
		\begin{array}{c}
			\nu_j\\
			e_j\\
		\end{array} \right)_L \sim (1,2,1,-1)~, ~ 
        \Psi_{jR} = \left(
		\begin{array}{c}
			\nu_j\\
			e_j\\
		\end{array} \right)_R \sim (1,1,2,-1)~, ~ 
    \end{split}
\end{equation*}
and the scalar sector only contains the two doublet fields
\begin{equation*}
    \chi_L \sim (1,2,1,1)~,\indent \chi_R \sim (1,1,2,1) ~.
\end{equation*}
Without a bi-doublet scalar, fermion mass terms can only be realized if four species of massive vector-like fermions are included whose gauge charges are given by 
\begin{equation*}
    T_k \sim(3,1,1,4/3)~,~ 
     B_k \sim(3,1,1,-2/3)~,~ 
      N_k \sim(1,1,1,0)~,~ 
       E_k \sim(1,1,1,-2)~,
\end{equation*}
and $k = 1,2$ is a generation index. Their electric charge can be inferred from
\begin{equation}
Q= T_L^3 + T_R^3 + \frac{X}{2}~,
\end{equation}
where $T_L^3~(T_R^3)$ gives the third component of $SU(2)_L~ (SU(2)_R)$ isospin.

To realize a double seesaw texture for fermion mass matrices, as discussed before, we propose to extend this classical setup by an extra abelian gauge symmetry, such that the gauge group becomes $\mathcal{G} = \mathcal{G}_{\mathrm{LRSM}}\times U(1)_{X'} $. While all particles mentioned so far are neutral with respect $U(1)_{X'}$, we consider one generation of fermions  that are vector-like with respect to $\mathcal{G}_{\mathrm{LRSM}}$ but are chiral under $U(1)_{X'}$:
\begin{equation*}
    \begin{split}
         T'_{L/R} &\sim(3,1,1,4/3, \mp\alpha)~,~ 
     B'_{L/R} \sim(3,1,1,-2/3,\pm \alpha)~,~\\
      N'_{L/R}&\sim(1,1,1,0, \pm\alpha)~,~
       E'_{L/R} \sim(1,1,1,-2, \mp\alpha)~.
    \end{split}
\end{equation*}
The anomaly-free charge assignment is arranged in such a way that explicit mass terms are forbidden for the new fermion species. Besides, we include a complex scalar singlet 
\begin{equation*}
    \eta \sim (1,1,1,0,\alpha)
\end{equation*}
that carries $U(1)_{X'}$ charge and is responsible for spontaneously breaking $U(1)_{X'}$. A summary of the particle spectrum together with assigned gauge charges can be found in Table~\ref{Tab.: particle contentII}.
\begin{table}[b!]
	\centering
	\resizebox{0.9\textwidth}{!}{%
	\begin{tabular}{|c|c|c|c|c|c|c|}
		\hline \hline
		\bf Type &\bf Particle   & \bf{$\mathbf{SU(3)_C}$}  & \bf{$\mathbf{SU(2)_R}$} &\bf{$\mathbf{SU(2)_L}$} &\bf{$\mathbf{U(1)_{X}}$}&$\mathbf{U(1)_{X'}}$\\
		\hline \hline
		\rowcolor{lightgray} 
		\multirow{2}{*}{\bf Quarks} &$Q_{jL} = \left(
		\begin{array}{c}
			u_j\\
			d_j\\
		\end{array}
		\right)_L$  & \bf{3} &  \bf{1} &  \bf{2}  & 1/3& 0\\
		\rowcolor{lightgray}
		&$Q_{jR} = \left(
		\begin{array}{c}
			u_j\\
			d_j\\
		\end{array}
		\right)_R $  & \bf{3} &  \bf{2} &  \bf{1}  & 1/3&0\\
		\hline
		\rowcolor{lightblue} 
		\multirow{2}{*}{\bf Leptons} &$\Psi_{jL} = \left(
		\begin{array}{c}
			\nu_j\\
			e_j\\
		\end{array}
		\right)_L$  & \bf{1} &  \bf{1} &  \bf{2}   & -1&0\\
		\rowcolor{lightblue}
		&$\Psi_{jR} = \left(
		\begin{array}{c}
			\nu_j\\
			e_j\\
		\end{array}
		\right)_R$  & \bf{1} &  \bf{2} &  \bf{1}   & -1&0\\
		\hline \hline
		\rowcolor{lightgreen} 
		 &$T_{1\, L/R}$ , $T_{2 \,L/ R}$ , $T'_{L/ R}$ & \bf{3} &  \bf{1} &  \bf{1}  & 4/3&$\{0,0,\mp \alpha\}$\\
		\rowcolor{lightgreen}
		\multirow{1}{*}{\textbf{BSM Fermions}} &$B_{1 \, L/R}$ , $B_{2 \,L/ R}$ , $B'_{L/ R}$ & \bf{3} &  \bf{1} &   \bf{1}  &-2/3&$\{0,0,\pm \alpha\}$\\
		\rowcolor{lightgreen}
		&$N_{1 \, L/R}$ , $N_{2 \, L/R}$ , $N'_{L/R}$ & \bf{1} &  \bf{1} &   \bf{1}  & 0&$\{0,0,\pm \alpha\}$\\
		\rowcolor{lightgreen}
		&$E_{1 \, L/R}$ , $E_{2 \, L/R}$ , $E'_{L/R}$ & \bf{1} &  \bf{1} &   \bf{1}  & -2&$\{0,0,\mp \alpha\}$\\
			\hline\hline
		\rowcolor{lightyellow} 
		\multirow{2}{*}{\textbf{Scalars}} &$\chi_L = \left(
		\begin{array}{c}
			\chi_L^{+}\\
			\chi_L^{0}\\
		\end{array}
		\right)$  & \bf{1} &  \bf{1} &  \bf{2}   & 1&0\\
		\rowcolor{lightyellow}
		&$\chi_R = \left(
		\begin{array}{c}
			\chi_R^{+}\\
			\chi_R^{0}\\
		\end{array}
		\right)$  & \bf{1} &  \bf{2} &  \bf{1}   & 1&0\\
		\rowcolor{lightyellow}
		&$\eta$ & \bf{1} &  \bf{1} & \bf{1} &$0$& $\alpha$\\
		\hline \hline
	\end{tabular}
	}
	\caption{Summary of the particle spectrum, where $j = 1,2,3$ labels the SM families, and we consider $m=2$ generations of vector-like fermions $(T,~ B, ~N,~ E)$ together with $n=1$ generation of primed fermions $(T',~ B', ~N',~ E')$.}
	\label{Tab.: particle contentII}
\end{table}

The scalar potential is given by
\begin{equation}
\label{Eq.:Scalar Potential}
\begin{split}
    V(\chi_L, \chi_R, \eta) = &\mu_1^2 \chi_L^\dagger \chi_L + \mu_2^2 \chi_R^\dagger \chi_R   + \lambda_{1L} (\chi_L^\dagger \chi_L)^2 + \lambda_{1R}(\chi_R^\dagger \chi_R)^2 + \lambda_2 (\chi_L^\dagger \chi_L) (\chi_R^\dagger \chi_R)\\
    +&\mu_\eta^2 \eta^\dagger \eta+\lambda_{3} (\eta^\dagger \eta)^2+
    \lambda_{4L}  (\eta^\dagger \eta) (\chi_L^\dagger \chi_L) + \lambda_{4R}  (\eta^\dagger \eta)
    (\chi_R^\dagger \chi_R) 
    \end{split}~.
\end{equation}
In a parity symmetric scenario,  we can identify  $\lambda_1 \equiv\lambda_{1L} = \lambda_{1R}$, $\lambda_4 \equiv\lambda_{4L} = \lambda_{4R}$
and  $\mu_1 = \mu_2$. Allowing $\mu_1 \neq \mu_2$ breaks parity softly. If the scalar fields acquire VEVs, the gauge symmetry is spontaneously broken. We can choose all the VEVs to be real by making gauge rotations. We assume the breaking scale of $U(1)_{X'}$ to be larger than that of $SU(2)_R$ and $SU(2)_L$. Therefore, symmetry breaking occurs in three steps
\begin{equation}
\begin{split}
   SU(3)_C \times SU(2)_R\times SU(2)_L \times U(1)_{X}\times U(1)_{X'} &\overset{\langle \eta \rangle}{\longrightarrow }SU(3)_C \times SU(2)_R\times SU(2)_L \times U(1)_{X} \\
   &\overset{\langle \chi_R \rangle}{\longrightarrow } SU(3)_C \times SU(2)_L \times U(1)_{Y} \\
    &\overset{\langle \chi_L \rangle}{\longrightarrow } SU(3)_C \times U(1)_{EM} ~.
   \end{split}
\end{equation}
We denote the VeVs of the three scalars by
\begin{equation}
	\langle\chi_L \rangle  = \dfrac{1}{\sqrt{2}}\left(
	\begin{array}{c}
		0\\
		v_{L}\\
	\end{array}
	\right) \, ,\indent
	\langle\chi_R\rangle  = \dfrac{1}{\sqrt{2}}\left(
	\begin{array}{c}
		0\\
		v_{R}\\
	\end{array}
	\right) \, ,\indent
	 \langle\eta \rangle  = \dfrac{v_\eta}{ \sqrt{2}}  \, ,
\end{equation}
where $v_L = 246 $ GeV denotes the electro-weak scale. The physical scalar spectrum of our model consists of three neutral scalars.
For simplicity, we consider $\lambda_4 \ll 1$. Then, the scalar $\eta $ decouples from the scalar mass spectrum and picks up a mass
$$
M_\eta^2 = 2 \lambda_3 v_\eta^2 ~.
$$
Diagonalizing the remaining two by two scalar mass matrix in the limit $v_L\ll v_R$ yields the mass eigenvalues
\begin{align}
    M_h^2 &\simeq \left(2 \lambda_1 - \dfrac{\lambda_2^2}{2 \lambda_1} \right) v_L^2 ~ ,\\
     M_H^2 &\simeq 2 \lambda_1 v_R^2 ~ ,
\end{align}
with a mixing angle $\tan(2 \xi) = \lambda_2 v_L v_R/\left(\lambda_1 (v_R^2-v_L^2) \right)$ (see also Appendix~A). The lighter of these two states can be identified with the SM Higgs boson. Since the bi-doublet is absent in our model, the charged gauge bosons $W^\pm_L = \left(W^2_L \mp i W^1_L\right)/\sqrt{2} $ and $W^\pm_R = \left(W^2_R \mp i W^1_R\right)/\sqrt{2}$ do not mix at tree-level and have the following masses
\begin{equation}
\label{Eq.: W mass}
	M^2_{W_L} = \dfrac{g_L^2 v_L^2}{4} ~, \indent 	M^2_{W_R} = \dfrac{g_R^2 v_R^2}{4} ~.
\end{equation}
Mixing between them only arises at the one-loop level, which will be a key ingredient for our mechanism to work. In total, our model has four neutral gauge bosons $(W^3_L, W^3_R, B, X)$ of which $X$ decouples from the spectrum and gets a mass
\begin{equation}
    M_{X}^2 = \dfrac{g_x^2 v_\eta^2 \alpha^2}{4} \, .
\end{equation}
Mixing between the remaining three gauge bosons results in a massless photon field $A_\mu$ and  two massive eigenstates $(Z_\mu, Z'_\mu)$ whose masses  are given by (see also Appendix A)
\begin{equation}
	M_{Z, Z'}^2 = \dfrac{1}{2} \left(M_{LL}^2+ M_{RR}^2 \pm (M_{LL}^2- M_{RR}^2)\sqrt{1+\tan^2(2\zeta)}\right) ~,
\end{equation}
where we defined
\begin{equation}
	\begin{split}
		M_{LL}^2 &= \dfrac{e^2 v_L^2}{4 s_w^2 c_w^2} ~,\\
		M_{RL}^2 &=   \dfrac{e^2 v_L^2}{4  c_w^2 \sqrt{c_w^2-s_w^2}} ~, \\
		M_{RR}^2 &=  \dfrac{e^2}{4 \left(c_w^2-s_w^2\right)}\left(\dfrac{c_w^2 v_R^2}{s_w^2}+ \dfrac{s_w^2 v_L^2}{c_w^2}\right) ~, \\
	\end{split}
\end{equation}
with the weak mixing angle $\sin(\theta_w) \equiv s_w$ and $\tan(2\zeta) = 2 M_{LR}^2/(M_{LL}^2- M_{RR}^2)$. For the case $g_R = g_L \simeq 0.65$, $g' = 0.428$, $\sin^2(\theta_w) = 0.231$ and a breaking scale of $v_R = 5$ TeV the mixing angle is $\zeta = 6.9 \times 10^{-4} $ and hence mixing effects will be hugely suppressed.

Finally, we will turn our interest to the possible Yukawa couplings. As a consequence of the charge assignments, the SM quark doublets can only couple to the vector-like quark singlets $T$ and $B$. In addition, there is a coupling between the vector-like quarks $T$ and $B$ and their non-vector-like partners $T'$ and $B'$ via the scalar singlet $\eta$.
Hence, the Lagrangian in the quark sector is given by
\begin{equation}
\label{Eq.: Quark Yukawa Couplings}
	\begin{split}
	\mathcal{L}_{Yuk} = &-y^q_a \overline{Q}_L \tilde{\chi_L} T_R -y^q_b  \overline{Q}_R \tilde{\chi_R} T_L \\
	&-y^q_c \overline{Q}_L \chi_L B_R -y^q_d  \overline{Q}_R \chi_R B_L \\
	&-y_1^q \overline{T}_L \eta^\dagger T'_R - y_2^q \overline{T}_R \eta T'_L \\
	&-y_3^q \overline{B}_L \eta B'_R - y_4^q \overline{B}_R \eta^\dagger B'_L
	 +~h.c.~,
	\end{split}
\end{equation}
where the $y^q$ indicates Yukawa coupling matrices.
On top of that, the vector-like quarks have explicit masses
\begin{equation}
\mathcal{L}_{VLF} =-M_T \overline{T}_L T_R -M_B \overline{B}_L B_R +~h.c. ~.
\end{equation}
with $M_T$ and $M_B$ being matrices of dimension $m \times m$. 

While the Yukawa couplings for charged leptons are quite similar to those of quarks, the vector-like neutrinos $N_{L,R}$ carry no $U(1)$ charges and allow additional lepton number violating (LNV) terms. We discuss these in more detail in Section~\ref{Sec.: Neutrino Masses} and only keep the lepton number conserving terms for now. Then, the Yukawa Lagrangian is given by
\begin{equation}
\label{Eq.: Yukawa Lagrangian Leptons}
	\begin{split}
		\mathcal{L}_{Yuk} = &-y^\ell_a \overline{\Psi}_L \tilde{\chi}_L N_R -y^\ell_b  \overline{\Psi}_R \tilde{\chi}_R N_L \\
		&-y^\ell_c \overline{\Psi}_L \chi_L E_R -y^\ell_d  \overline{\Psi}_R \chi_R E_L \\
		&-y_1^\ell \overline{N}_L \eta N'_R - y_2^\ell \overline{N}_R \eta^\dagger N'_L 
	 \\
		&-y_3^\ell \overline{E}_L \eta^\dagger E'_R - y_4^\ell \overline{E}_R \eta E_L^{'}
		+~h.c. ~,
	\end{split}
\end{equation}
where the $y^\ell$ denote the various Yukawa coupling matrices, and the explicit mass terms are given by
\begin{equation}
\label{Eq.: Explicit masses leptons}
	\mathcal{L}_{VLF} =-M_N \overline{N}_L N_R -M_E \overline{E}_L E_R  +  ~h.c. ~.
\end{equation}
In the following, we demand the model to be symmetric under  parity symmetry, which maps the following fields
\begin{equation}
    \begin{split}
    Q_L \leftrightarrow Q_R~, \indent  \Psi_L \leftrightarrow \Psi_R~, \indent T^{(')}_L \leftrightarrow T^{(')}_R~, \indent B^{(')}_L \leftrightarrow B^{(')}_R~,\\  \indent N^{(')}_L \leftrightarrow N^{(')}_R~, \indent E^{(')}_L \leftrightarrow E^{(')}_R~,
    \indent \chi_L \leftrightarrow \chi_R~, \indent \eta \leftrightarrow \eta^{\dagger}~. 
    \end{split}
\end{equation}
The number of free parameters is now  reduced significantly as we impose parity and identify
\begin{equation}
    \begin{split}
        y_a^q = y_b^q~,~~  y_c^q = y_d^q~,~~ y_1^q = y_2^q~,  ~~ y_3^q = y_4^q~,
    \end{split}
\end{equation}
and similarly in the lepton sector
\begin{equation}
    y_a^\ell = y_b^\ell~, ~~y_c^\ell = y_d^\ell~, ~~y_1^\ell = y_2^\ell~ , ~~y_3^\ell = y_4^\ell~.
\end{equation}
For vector-like mass matrices, this implies 
\begin{equation}
    M_T = M_T^\dagger~, \indent  M_B = M_B^\dagger~, \indent M_N = M_N^\dagger~, \indent M_E = M_E^\dagger~.
\end{equation}
For simplicity, we assume these matrices will be diagonal and real in the remainder of this paper. This can be done on a suitable basis and without loss of generality.
\section{Quark masses}\label{SEC-03}
When the scalar fields acquire VEVs, the Lagrangian in Eq.~(\ref{Eq.: Quark Yukawa Couplings}) gives rise to a mass matrix for the up-type quarks. For ease of reading, we drop the superscripts $q$ in the Yukawa couplings in this section. We consider now a specific case with  $m=2$ generations of massive vector-like fermions and $n=1$ generation of massless fermions that are vector-like with respect to the gauge group  $\mathcal{G}_{\mathrm{LRSM}}$ but chiral under $U(1)_{X'}$.
At tree-level, the $6 \times 6 $ up quark mass matrix in the basis $(u_1, u_2, u_3, T_1,T_2, T')$ takes the form
\begin{equation}
\label{Eq.: mass matrix tree level}
    \mathcal{M}_u^{(0)} = \left(
    \begin{array}{ccc}
        \mathbf{0}& v_L y_a/\sqrt{2} & \mathbf{0} \\
        v_R  y_a^\dagger/\sqrt{2} &M_T & v_\eta y_1/\sqrt{2}  \\
        \mathbf{0} & v_\eta y_1^\dagger /\sqrt{2}&\mathbf{0}
    \end{array}
    \right)~,
\end{equation}
where $y_a$ and $y_1$ are the $3 \times 2$ and $2 \times 1$ Yukawa matrices, specified in Eq.~(\ref{Eq.: Quark Yukawa Couplings}). At this level, all zero entries are a consequence of the assigned gauge charges and particle content.
The shown mass matrix has rank four in total. Separating into three light and three heavy states yields that $M^{\mathrm{heavy}}$ has full rank three while the light eigenvalue matrix $M^{\mathrm{light}}$ has rank one. In other words, the two massive vector-like fermions $T_1, T_2$ give tree-level mass to the fermion $T'$ and to the top quark. All other remaining particles are massless at this stage.

To further analyze the matrix with regard to its eigenvalues, we reshape it to the familiar form of a type-I seesaw matrix by defining
\begin{equation}
    \mathcal{M}_S = \left(
    \begin{array}{cc}
        M_T & v_\eta y_1/\sqrt{2} \\
        v_\eta y_1^\dagger/\sqrt{2}  &\mathbf{0} 
    \end{array}
    \right)
\end{equation}
 and $ Y_a \equiv \left( y_a,~ \mathbf{0}\right)$. Thus, Eq.(\ref{Eq.: mass matrix tree level}) can be rewritten as \begin{equation}
     \mathcal{M}_u^{(0)} = \left(
    \begin{array}{cc}
        \mathbf{0}& v_L Y_a /\sqrt{2} \\
        v_R Y_a^\dagger /\sqrt{2}&\mathcal{M}_S 
    \end{array}
    \right) ~.
 \end{equation}
In the limit $ v_\eta , M_T > v_R > v_L $, the light and heavy eigenvalues are approximately given by 
\begin{equation}
    \begin{split}
        M^{\mathrm{light}} &\simeq - \dfrac{v_L v_R}{2} Y_a \mathcal{M}_S^{-1} Y_a^\dagger ~, \\
        M^{\mathrm{heavy}} &\simeq \mathcal{M}_S ~.
    \end{split}
\end{equation}
Using the definition of $Y_a $ we can rewrite the expression $Y_a \mathcal{M}_S^{-1} Y_a^\dagger = y_a (\mathcal{M}_S^{-1})_{1,1} y_a^\dagger$ which shows that we are only interested in the (1,1)-block of $\mathcal{M}_S^{-1}$.
Inverting the block diagonal matrix $\mathcal{M}_S$ yields 
\begin{equation}
    \mathcal{M}_S^{-1} = \left(
    \begin{array}{cc}
        M_T^{-1}- M_T^{-1}y_1 \left(y_1^\dagger  M_T^{-1}y_1  \right)^{-1} y_1^\dagger  M_T^{-1} &  \sqrt{2}v_\eta^{-1} M_T^{-1} y_1  \left( y_1^\dagger  M_T^{-1}y_1 \right)^{-1} \\
         \sqrt{2} v_\eta^{-1}\left( y_1^\dagger  M_T^{-1}y_1 \right)^{-1} y_1^\dagger M_T^{-1}& - 2 v_\eta^{-2}\left( y_1^\dagger  M_T^{-1}y_1 \right)^{-1} 
    \end{array}
    \right) ~.
\end{equation}
Hence, we find
\begin{equation}
    \begin{split}
    M^{\mathrm{light}} &\simeq -\dfrac{v_L v_R}{2} y_a \left(  M_T^{-1}- M_T^{-1}y_1 \left(y_1^\dagger  M_T^{-1}y_1  \right)^{-1} y_1^\dagger  M_T^{-1}\right) y_a^\dagger \\
    & = -\dfrac{v_L v_R}{2} y_a M_T^{-1} y_a^\dagger+ \dfrac{v_L v_R}{2} y_a  \left(M_T^{-1}y_1 \left(y_1^\dagger  M_T^{-1}y_1  \right)^{-1} y_1^\dagger  M_T^{-1}\right) y_a^\dagger
\\
M^{\mathrm{heavy}} &\simeq \mathcal{M}_S
    \end{split}
\end{equation}
The eigenvectors corresponding to the zero mass states at the tree-level are shown explicitly in the next section.

\begin{figure}[t]
	\centering
	\includegraphics[scale=0.23]{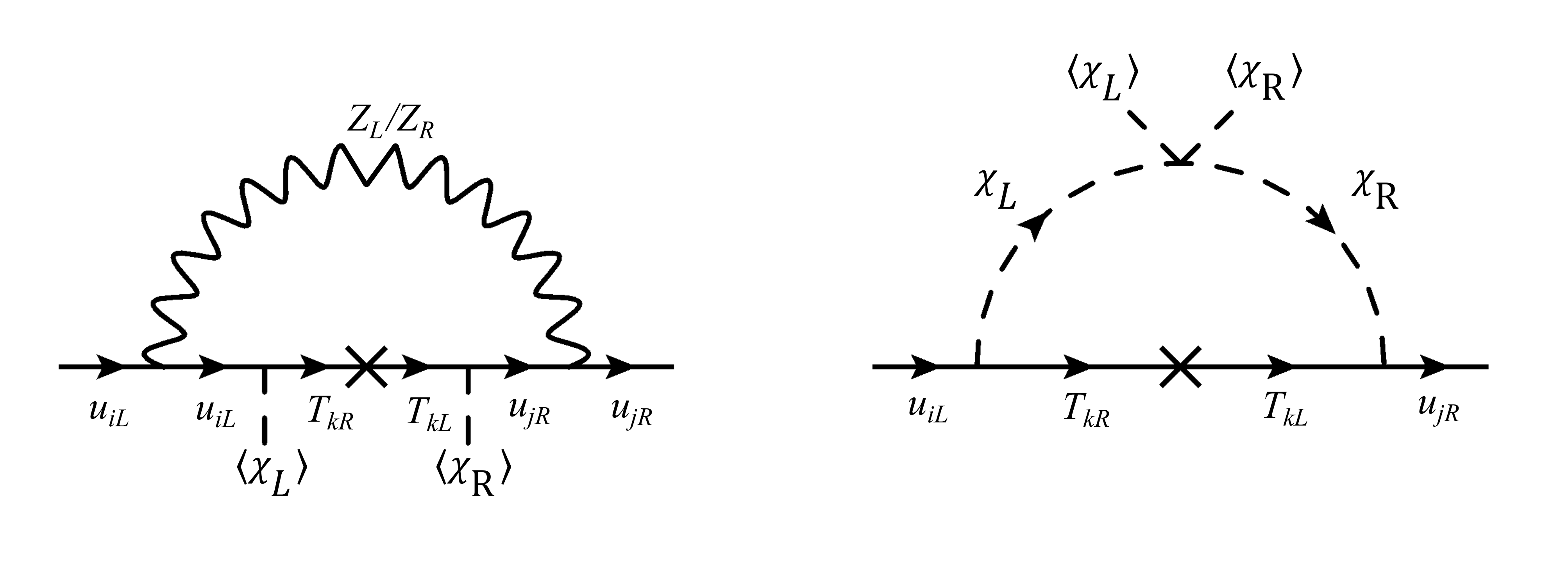}
	\caption{One-loop corrections to the up-type quark mass matrix, which generates the charm quark mass.}
	\label{Fig.: Relevant one-loop contributions}
\end{figure}

At one loop, corrections to the mass matrix populate formerly zero entries and give mass to further SM fermions by enhancing the matrix rank. The most relevant one-loop diagrams contributing to the upper left zero entry of the mass matrix can be seen in Fig.~\ref{Fig.: Relevant one-loop contributions}. We denote their contribution by $\delta M_u^{(1)}$. Another contribution, $\delta \tilde{M}_u^{(1)}$,  arises from $X$ and $\eta$ exchange. On top of that, mixed $\chi_L$ ($\chi_R$) and $\eta$ exchange populate the formerly zero entry in the upper right (lower left) entry of the mass matrix. A summary of the one-loop graphs is given in Fig.~\ref{Fig.: Systematic of one-loop contributions} and a more detailed discussion of the corresponding loop contributions can be found in Appendix~C. Overall, the one-loop corrected mass matrix is  given by
\begin{equation}
\label{Eq.: mass matrix 1-loop level}
    \mathcal{M}_u^{(1)} = \left(
    \begin{array}{ccc}
        \delta M_u^{(1)}& v_L y_a/ \sqrt{2} & \delta v_L \\
        v_R y_a^\dagger/ \sqrt{2} &M_T & v_\eta y_1 / \sqrt{2} \\
        \delta^\dagger v_R & v_\eta y_1^\dagger / \sqrt{2}&\delta \tilde{M}_u^{(1)}
    \end{array}
    \right) ~.
\end{equation}
In our scenario, $\delta M_u^{(1)} \sim y_a y_a^\dagger$, $\delta \tilde{M}_u^{(1)} \sim y_1^\dagger y_1 $ and $\delta \sim y_a y_1$. When added to the tree-level mass matrix, the rank increases further by one, such that the charm quark becomes massive at one-loop. 
In the seesaw limit, the corrected eigenvalues are 
\begin{equation}
    \begin{split}
        M^{\mathrm{light}} &\simeq \delta M_u^{(1)}  - \dfrac{v_L v_R}{2} Y_a^{(1)} \left(\mathcal{M}_S^{(1)}\right)^{-1} Y_a^{(1)\dagger} \\
        M^{\mathrm{heavy}} &\simeq \mathcal{M}_S^{(1)} ~,
    \end{split}
\end{equation}
where $\mathcal{M}_S^{(1)} $ is given by
\begin{equation}
\label{Eq.: One-loop corrected M_S}
    \mathcal{M}_S^{(1)} =  \left(
    \begin{array}{cc}
        M_T & v_\eta y_1 / \sqrt{2} \\
        v_\eta y_1^\dagger/ \sqrt{2}  &\delta \tilde{M}_u^{(1)}
    \end{array}
    \right) ~,
\end{equation}
and we defined $Y_a^{(1)} \equiv \left( y_a,~ \delta \sqrt{2}\right)$.
While $\delta \tilde{M}_u^{(1)}$ mainly adds to the mass of the exotic fermion $T'$, the more interesting part is the magnitude of  $\delta M_u^{(1)}$ which should be at the order of the charm quark mass. An estimate for the leading contribution to the left diagram in Fig.~\ref{Fig.: Relevant one-loop contributions} yields 
\begin{equation}
\begin{split}
    m_c \sim \delta M_u^{(1)} &\simeq \dfrac{g^2   }{16 \pi^2} v_L v_R y_a \dfrac{M_T}{M_T^2-M_Z^2} y_a^\dagger \\&\simeq \dfrac{g^2   }{16 \pi^2} v_L v_R  y_a M_T^{-1} y_a^\dagger 
    ~,
\end{split}
\end{equation}
where we assume $M_T > v_R > v_L $. Evidently, we find a $1/(16\pi^2)$ suppression with respect to the third-generation mass. The explicit results for the loop correction that is used for the numeric fit are given in  Appendix~C.

\begin{figure}[t!]
    \begin{subfigure}{0.48 \textwidth}
    \vfill
    \vfill
	    \centering
	    \includegraphics[scale=0.1]{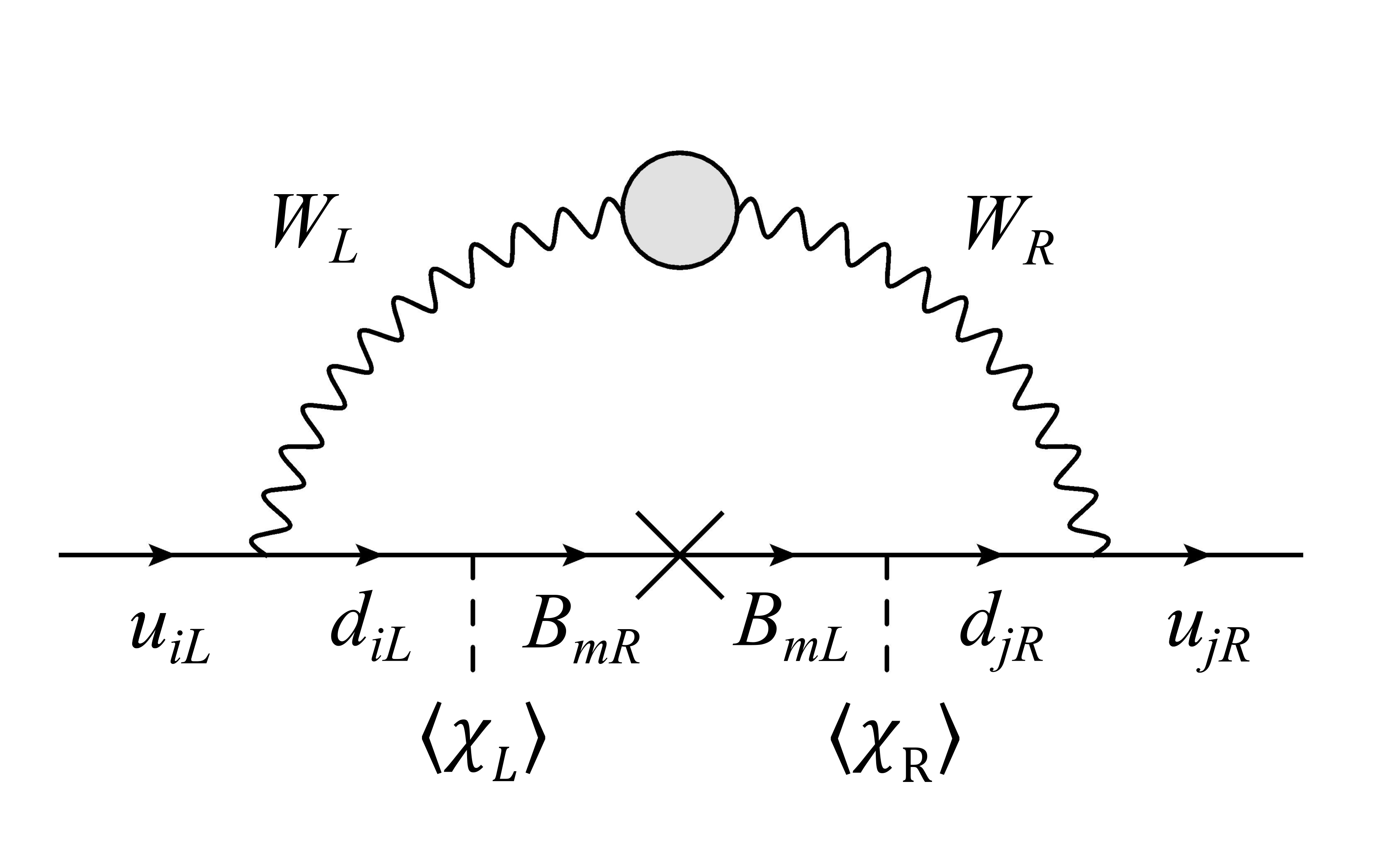}
	    \vphantom{T}
	    \subcaption{}
    	\label{Fig.: Two-Loop W exchange}
	\end{subfigure}
	\hfill
	\begin{subfigure}{0.48 \textwidth}
	    \centering
	    \includegraphics[scale=0.088]{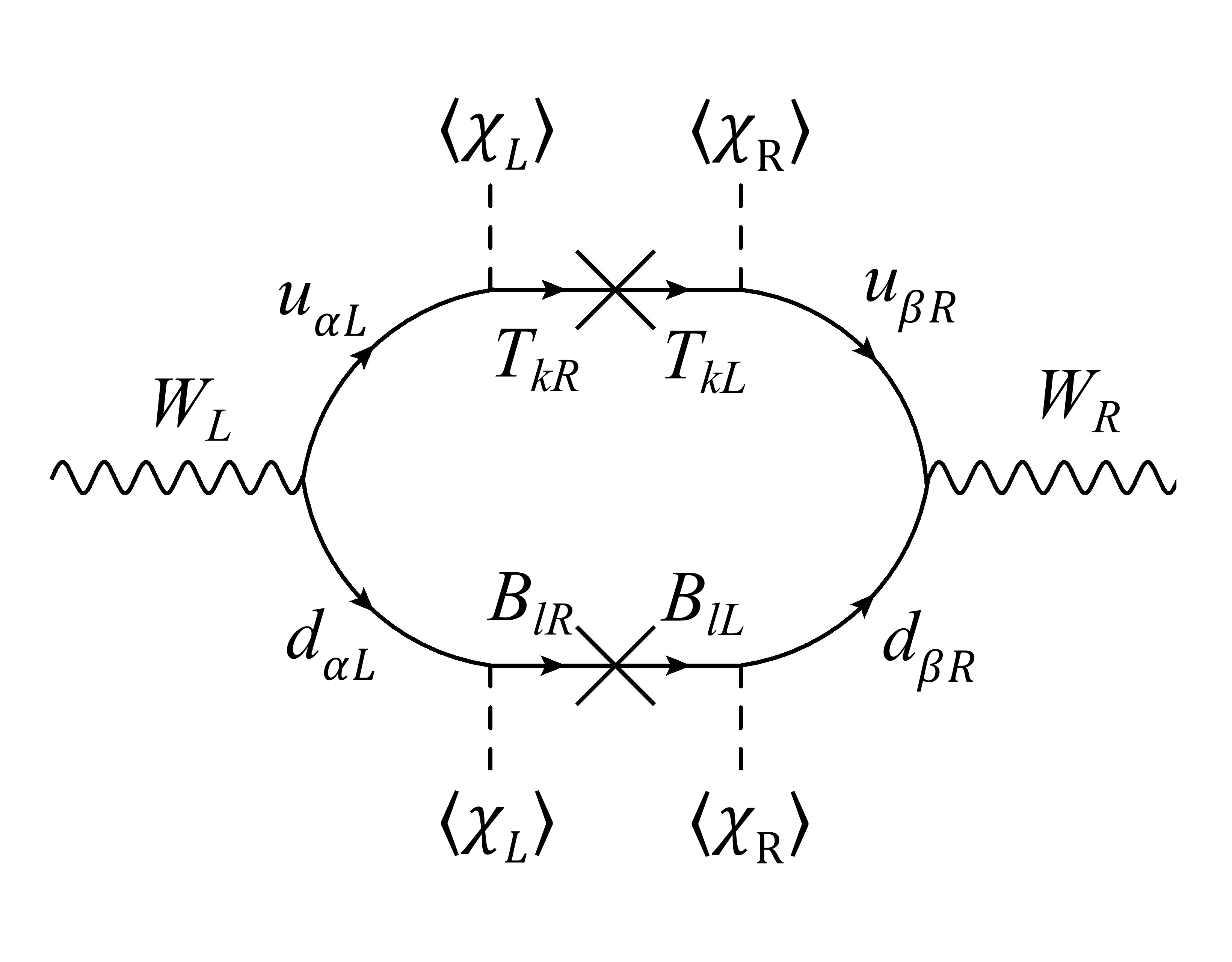}
	    \subcaption{}
	    \label{Fig.: WLWR_mixing}
	\end{subfigure}
 \caption{(a) Two-loop radiative correction to the quark mass matrix that generates the up quark mass. The grey blob indicates mixing of $W_L-W_R$ gauge bosons at one-loop~\cite{Chang:1986bp} and is expanded in (b). }
\end{figure} 
At two-loop order, there are further corrections, among which the most important one is given by the  $W_L-W_R$ exchange diagram shown in Fig.~\ref{Fig.: Two-Loop W exchange}. Note that the mixing of $W_L-W_R$ only happens at one-loop level due to the scalar particle content of our model (see Fig.~\ref{Fig.: WLWR_mixing}). The diagram in Fig.~\ref{Fig.: Two-Loop W exchange} is proportional to the matrix $\delta M_u^{(2)} \sim y_c y_c^\dagger$ where $y_c $ is the down-type Yukawa coupling and in general $y_a$ and $y_c$ can be linear independent. Together, we get
\begin{equation}
\label{Eq.: mass matrix 2-loop level}
    \mathcal{M}_u^{(2)} = \left(
    \begin{array}{ccc}
        \delta M_u^{(1)} +\delta M_u^{(2)}& v_L y_a/ \sqrt{2} & \delta v_L \\
        v_R y_a^\dagger/ \sqrt{2} &M_T & v_\eta y_1 / \sqrt{2} \\
        \delta v_R & v_\eta y_1^\dagger / \sqrt{2}&\delta \Tilde{M}_u^{(1)}
    \end{array}
    \right) ~,
\end{equation}
where $\mathcal{M}_u^{(2)}$ is now a rank six matrix. Thus, first-generation masses are only generated at two-loop order.
The full expression for the two-loop radiative correction to the mass matrix is derived in  Appendix~C. If all vector-like masses are roughly at the same scale $M^2_{T} \simeq M^2_{B}$, the approximate two-loop contribution that generates the up quark mass is given by
\begin{equation}
\label{Eq.: 2-loop approximation}
    \begin{split}
        m_u\sim \delta M_u^{(2)} \simeq
 \dfrac{N_c}{(16 \pi^2)^2}  \dfrac{g^4 v^3_L v^3_R M_{B} }{M_{T} M_{B} M_{W_L}^2 M_{W_R}^2} ~,
    \end{split}
\end{equation}
for $\mathcal{O}(1)$ Yukawa couplings. This clearly shows the suppression of first-generation fermion masses compared to the other generations. A quick estimate of the size of the contribution reveals that the scale of new physics should roughly fulfill the relation $v_R/M_T \simeq 10^{-1}$.  

\section{ Explicit demonstration of the stepwise rank increase}
\label{Sec.: eigenstates}
In this section, we show explicitly the working of the rank mechanism in our model i.e. we show that there are two massless states at tree-level, which pick up hierarchical masses step-wise at  one- and two-loops. 
For that purpose, we calculate the eigenstates of the  tree-level fermion mass matrix. Let us just focus on the up quark sector of the mass matrix in Eq.~(\ref{Eq.: mass matrix tree level}). We can work in a basis where $T_{1,2}$ mass matrix is diagonal without loss of generality. We can, then, do a unitary rotation of the $(Q_1, Q_2, Q_3)$ prior to gauge symmetry breaking and go to a basis so that we have the Yukawa coupling matrix $y^q_a$  written as follows:
\begin{equation}
\label{Eq.: massive quark states tree level}
y^q_a~=~\left(\begin{array}{cc} 0 & 0\\0 & b\\a&c\end{array}\right);~~
y^q_1~=~\left(\begin{array}{c} n_1\\n_2\end{array}\right)~.
\end{equation}
We do not lose any generality by this choice since this is a unitary transformation. 
The tree-level massive eigenstates of $\mathcal{M}_u^{(0)}$ then are
\begin{equation}
|0\rangle = N_0[a|u_3\rangle +n_1|T'\rangle];~~|3\rangle =N_1[ b|u_2\rangle +c|u_3\rangle+n_2|T'\rangle]
\end{equation}
with the $|0\rangle$ pairing up with $|T_1\rangle$ and $|3\rangle$ pairing up with $|T_2\rangle$ to form the Dirac masses in the tree-level seesaw matrix. (The $N_a$ are the normalization factors.) This leads to two massive fermion states. They correspond to the massive vector-like fermion state and the top quark state, respectively. The top quark state is a linear combination of $u_2, u_3$ and $T'$ states pairing up with the opposite helicity of linear combination of $T_{1,2}$ and $u_{2,3}$ states after seesaw diagonalization of the tree-level masses of the top quark and the heavy states. We get $m_t\simeq \frac{y^2_a v_Lv_R}{M} $, with $M$ being a function of $M_{T1,T2}$. The states $|0\rangle$ and $|3\rangle$ are not orthogonal, but they define a plane in the four-dimensional space of $u_{1,2,3}$ and $T'$.

The states orthogonal to $|0\rangle$ and  $|3\rangle$ are massless and are given by
\begin{equation}
    |2\rangle_{L,R} ~=~N_2[(cn_1-an_2)|u_2\rangle -n_1b| u_3\rangle +ab|T'\rangle];~|1\rangle_{L,R} ~=~|u_1\rangle
\end{equation}

The states $|2\rangle$ and $|1\rangle$ correspond to the charm and up quark states, respectively. The latter two pick up mass only at the one- and two-loop level. One can check that in the basis we have chosen, none of the one-loop diagrams in Fig.~\ref{Fig.: Systematic of one-loop contributions} contribute to the first generation up quark mass. To see that only  $|2\rangle_{L,R}$ picks up mass at one-loop, note that  $|1\rangle_{L,R}$ does not have any Yukawa coupling to the heavy quarks (note the two zero entries in row one $y^q_a$). That allows a chiral $U(1)$ symmetry operating on this field in the Lagrangian except in its $W_{L,R}$ coupling. That allows it to have mass only at two loop since $W_L-W_R$ mixing is a one loop effect in this theory (see the two-loop graph in Fig.~\ref{Fig.: Two-Loop W exchange}).

The $y$ couplings in the up and down sector are not related. So, once we have used the freedom to rotate the up sector, we are not free to do the same in the down quark sector since the rotation we did was for the quark doublet $Q$ which contains both the up and down quarks. After symmetry breaking however, the up and down sectors become ``separate'' and we can rotate the two sectors independently. The down sector Yukawa matrices then take the same form as in Eq.~(\ref{Eq.: massive quark states tree level}). However, when we do that, there would be a relative ``tilt'' between the up and down quarks in the charged current $W^+_{L, R}$ interaction, which would give non-zero CKM mixings at the tree-level. As a result, when we use the two-loop $W_L-W_R$ mixing graph to calculate the up quark mass, the graph in Fig.~\ref{Fig.: Two-Loop W exchange} will connect the external up quark to the down quarks of the second and third generations, which are already massive at one-loop and tree-levels, respectively. This will then lead to a nonzero first-generation up quark mass at two loops. The above procedure can be applied to calculate the down quark masses at the one- and two-loop levels with similar conclusions.

Also, we note that the internal quark states contributing to two-loop first-generation up and down quark masses are the down and up quarks of higher generations, respectively. Since all second- and third-generation down quarks are lighter than the corresponding up quarks, this explains why the first-generation up quark is lighter than the down quark, resulting in an inverted hierarchy $m_u < m_d$.

We note further that, even though quark mass hierarchies can be explained by our method, the hierarchy in quark mixings cannot be due to the arbitrary tree-level up-down mixing ``tilt'' just mentioned. This is because the tilt depends on tree-level parameters and is arbitrary.  We wish to explore in a future publication whether adding a possible new discrete global symmetry to the model can eliminate this arbitrary tilt and make the mixing angles calculable. We note that the benchmark points in Table~\ref{Tab:yukawabenchmarks} are given in a different basis for quarks than this Section.

The above method can also be applied to the charged lepton sector and to calculate the neutrino Dirac masses. To understand the neutrino masses, however, we will need to give Majorana masses to the $N$ leptons and use the type I seesaw to understand small neutrino masses, as argued in Sec.~\ref{Sec.: Neutrino Masses}, Eq.~(\ref{Eq.: Neutrino mass Complete tree level}).

To conclude this section, we would like to note that the mass generation in the down-type sector works completely analogously with the obvious replacements of Yukawa couplings and vector-like mass matrices.

\section{Charged lepton masses}\label{SEC-04}
\begin{figure}[htb!]
	\centering
	\includegraphics[scale=0.22]{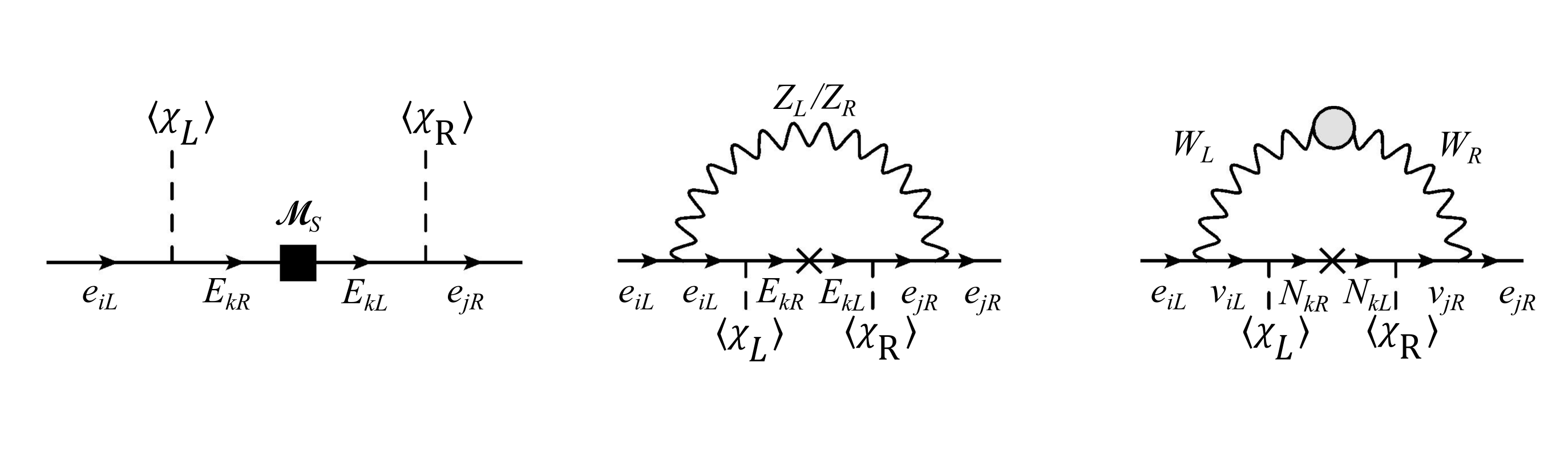}
\caption{Schematic of the charged lepton mass generation at the tree-level, one-loop, and two-loop.}
\label{Fig.: Scheme_ChargedLeptons }
\end{figure}

The generation of charged lepton masses proceeds very similarly to the quark sector (see Fig.~\ref{Fig.: Scheme_ChargedLeptons }). At the tree-level, the mass matrix is given by Eq.~(\ref{Eq.: mass matrix tree level}) with adequate replacements of the Yukawa couplings and vector-like masses. One-loop diagrams enhance the tree-level mass matrix rank by one and thereby generate the muon mass in a manner similar to the quark sector. The two-loop contribution that generates the electron mass originates from the insertion of the vector-like mass $M_N$ as shown in the last diagram in Fig.~\ref{Fig.: Scheme_ChargedLeptons }. Evidently, this diagram is proportional to the Dirac mass of neutrinos. If one assumes that there is no significant hierarchy in the vector-like masses of different fermion species $M_N \simeq M_E\simeq M_B \simeq M_T$ and with order one Yukawa couplings, the electron mass can be easily generated at the right scale. This gives rise to the question of whether neutrino masses can be correctly described in such a scenario, which we discuss in the next section. 

\section{Neutrino masses}
\label{Sec.: Neutrino Masses}
Without any lepton number violating operators in the theory, the neutrinos receive a Dirac mass in an analogous manner to the quarks. The neutrino Dirac mass matrix, including one and two-loop graphs is given by  
\begin{equation}
	\mathcal{M}^{(2)}_{\nu, \mathrm{Dirac} } = \left(
	\begin{array}{ccc}
		\delta M_\nu^{(1)}+ \delta M_\nu^{(2)}&  y_a v_L/\sqrt{2} &	\delta v_L \\
	y^{  \dagger}_a v_R/ \sqrt{2}&M_N&y_1 v_\eta/\sqrt{2}\\
	\delta^\dagger v_R& y^{ \dagger}_1 v_\eta/\sqrt{2} &\delta \tilde{M}_\nu^{(1)}
	\end{array}
	\right)~,
\end{equation}
which is a rank six matrix. Note that we drop the superscript $y^\ell$ in this section to improve readability. If $M_N$ is at the same scale as $M_E$ and $y_a \sim \mathcal{O}(1)$, this generates Dirac masses for the neutrinos that are of the order of the charged lepton masses. By applying a block matrix diagonalization, we find three light and three heavy neutrino states
\begin{equation}
    \begin{split}
        M_{\nu, \mathrm{Dirac}}^{\mathrm{light}} &\simeq (\delta M_\nu^{(1)}+\delta M_\nu^{(2)})  - \dfrac{v_L v_R}{2} Y_a \left(\mathcal{M}_S^{(1)}\right)^{-1} Y_a^{\dagger} \\
        M_{\nu, \mathrm{Dirac}}^{\mathrm{heavy}} &\simeq \mathcal{M}_S^{(1)} ~,
    \end{split}
\end{equation}
where $Y_a \equiv \left(y_a , \delta \sqrt{2}\right)$
and $\mathcal{M}_S^{(1)}$ is given by 
\begin{equation}
    \mathcal{M}_S^{(1)} = \left(
	\begin{array}{cc}
	M_N&y_1 v_\eta/\sqrt{2}\\
	 y^{ \dagger}_1 v_\eta/\sqrt{2} &\delta \tilde{M}_\nu^{(1)}
	\end{array}
	\right)~.
\end{equation}
Note that we did not include any two-loop corrections to $\mathcal{M}_S^{(1)}$, as this is already a full rank matrix, and two-loop corrections would only constitute minor changes. To comply with experimental bounds on neutrino masses, there needs to be a further suppression of $M_{\nu, \mathrm{Dirac}}^{\mathrm{light}}$ in order to obtain neutrino masses  $< \mathcal{O}(1 ~\mathrm{eV})$. 

If, at some scale, the theory gives rise to LNV operators, the most general neutral fermion sector Lagrangian becomes (using the notation where all fields are left-handed and fermion fields with superscript c denote the charge conjugate of right-handed fields and where parity symmetry transforms $F\to F^c$):
\begin{eqnarray}
{\cal L}_\nu=y_{a} {\Psi}^T\tilde{\chi}_LN^c+y_a^{'} {\Psi}^T \tilde{\chi}_L N + M_N N N^c + y_{1} N^c\eta N'+ M'_{L} N N  +y_1^{'} N\eta N' +\\\nonumber
y_{a} {\Psi^c}^T\tilde{\chi}_R N+y_a^{'} {\Psi^c}^T\tilde{\chi}_R N^c + M_N {N^c} N+y_{1} N \eta^* N^{c'}+M'_{R} N^c N^c +y_1^{'} N^c \eta N^{c'} +h.c.
\end{eqnarray}
Primed parameters denote LNV ones, whereas unprimed ones are lepton number conserving. Thus, the $M'$ terms are the LNV masses, and $y'$'s are LNV couplings. We let the soft LNV masses break parity in the lepton sector, i.e., $M'_{L}\neq M'_{R}$.

This Lagrangian, after spontaneous breaking, leads to the following mass matrix in the basis $(\nu, \nu^c, N_1, N^c_1, N_2, N^c_2, N' , N^{' c})$ (all fields left-handed):

\begin{eqnarray}
\label{Eq.: Neutrino mass Complete tree level}
{\cal M}_\nu^{(0)} =\left(\begin{array}{cccccccc} 0 & 0 & y^{'i1}_a \frac{v_L}{\sqrt{2}} & y^{i1}_a \frac{v_L}{\sqrt{2}} &y^{'i2}_a \frac{v_L}{\sqrt{2}} & y^{i2}_a \frac{v_L}{\sqrt{2}} & 0 & 0\\ 0 & 0 & y^{i1}_a \frac{v_R}{\sqrt{2}} &  y^{'i1}_a \frac{v_R}{\sqrt{2}} & y^{i2}_a  \frac{v_R}{\sqrt{2}} &  y^{'i2}_a \frac{v_R}{\sqrt{2}}& 0 & 0\\
y^{'i1}_a \frac{v_L}{\sqrt{2}} &  y^{i1}_a  \frac{v_R}{\sqrt{2}} & M'_{L,11} & M_{N1} & M'_{12} & 0 & y^{'1}_1\frac{v_\eta}{\sqrt{2}} & y^{1}_1\frac{v_\eta}{\sqrt{2}}\\ y^{i1}_a  \frac{v_L}{\sqrt{2}}  &y^{'i1}_a \frac{v_R}{\sqrt{2}}  & M_{N1} & M'_{R,11} &0 & M'_{12} &  y^{1}_1 \frac{v_\eta}{\sqrt{2}} & y^{'1}_1 \frac{v_\eta}{\sqrt{2}}\\
y^{'i2}_a \frac{v_L}{\sqrt{2}}& y^{i2}_a  \frac{v_R}{\sqrt{2}} &M'_{12}  & 0 & M'_{L,22} & M_{N2}  & y^{'2}_1 \frac{v_\eta}{\sqrt{2}} & y^{2}_1 \frac{v_\eta}{\sqrt{2}}\\ y^{i2}_a \frac{v_L}{\sqrt{2}}& y^{'i2}_a \frac{v_R}{\sqrt{2}}  & 0 &  M'_{12} & M_{N2} & M'_{R,22} & y_1^{2} \frac{v_\eta}{\sqrt{2}}& y^{'2}_1\frac{v_\eta}{\sqrt{2}}\\
0 & 0 & y^{'1}_1\frac{v_\eta}{\sqrt{2}} & y^{1}_1\frac{v_\eta}{\sqrt{2}}  &  y^{'2}_1\frac{v_\eta}{\sqrt{2}}& y^{2}_1\frac{v_\eta}{\sqrt{2}}  &0 & 0 \\ 0 & 0 & y^{1}_1\frac{v_\eta}{\sqrt{2}} & y^{'1}_1\frac{v_\eta}{\sqrt{2}}  &  y^{2}_1\frac{v_\eta}{\sqrt{2}} & y^{'2}_1\frac{v_\eta}{\sqrt{2}} &0 & 0 \end{array}\right)~,
\end{eqnarray}
where superscripts of the matrices $y_a,~ y_a^{'}$ and $y_1,~ y_1^{'}$ label the corresponding matrix entries and we only indicate couplings to one active neutrino generation.
We have to analyze the matrix in Eq.~(\ref{Eq.: Neutrino mass Complete tree level}) to see if we can have light neutrinos for some parameter regions of the model in order to satisfy neutrino oscillation data.
In its full form (where $i=1,2,3$), the matrix has dimension $12 \times 12$, and the derivation of analytic expressions for its eigenvalues is challenging due to the numerous parameters involved. However, as the matrices $M'_L$ and $M'_R$ both have rank two, we can anticipate that four neutrino states can be light in our scenario. A detailed treatment of the whole matrix, including loop corrections, necessitates a comprehensive parameter study, which is beyond the scope of this work. Instead, we analyze a simplified scenario of the tree-level mass matrix, where only the third-generation neutrino is considered ($i=3$). Our results are presented in the following section.
\begin{figure}[htb!]
	\centering
 $$\includegraphics[width=0.48\textwidth]{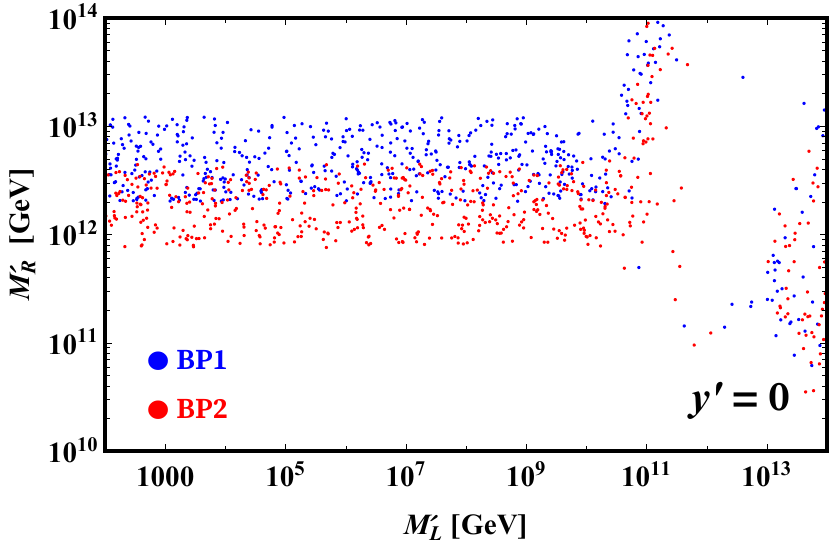}~~~~~
 \includegraphics[width=0.48\textwidth]{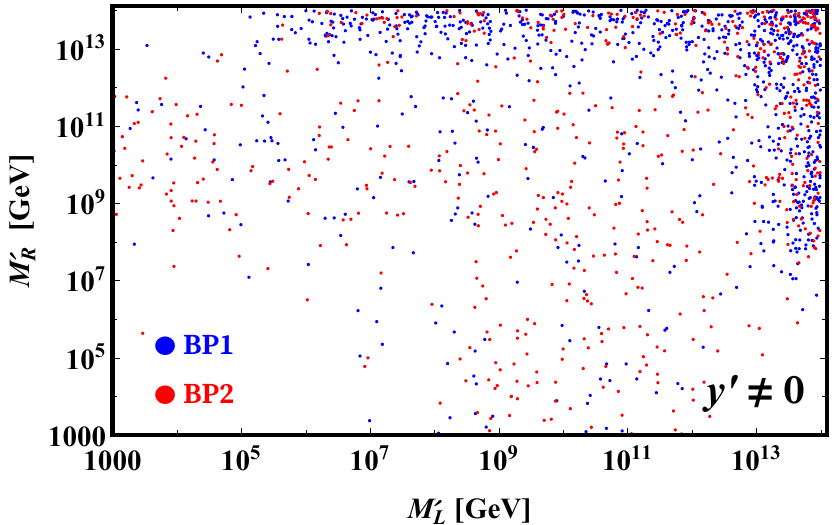} $$
\caption{Parameters in the $M'_L-M'_R$ plane which reproduces a light active neutrino with mass below $0.3$ eV at the tree-level. Lepton number conserving Yukawa couplings and masses for BP1 and BP2 are taken from Table~\ref{Tab:yukawabenchmarks}. For the parameter points in the left panel, all LNV couplings $y'$ are set to zero, whereas the right panel randomly scatters $y'$ couplings in a range $\{0.1,~1.0\}$.  }
\label{Fig.:neutrino scatter}
\end{figure}
\section{Numerical results}
\begin{table}[htb!]
	\centering 
	\resizebox{0.5\textwidth}{!}{%
		\begin{tabular}{|c|c|c|}
			\hline\hline
			\multirow{2}{*}{\begin{tabular}[c]{@{}c@{}}\bf{Yukawa Couplings}\\\end{tabular}} & \multicolumn{2}{c|}{\bf{Benchmark Points}}     \\ \cline{2-3} 
			&\bf{BP1}     & \bf{BP2}      \\ \hline \hline
			\rowcolor[HTML]{FFFFFF} 
			\multicolumn{1}{|c|}{\cellcolor[HTML]{FFFFFF}
				$y^{q}_a$     }                                                              &  $\left(
			\begin{array}{cc}
				0.616&  0.345\\ 0.543&0.742\\  2.291&  0.905\\
			\end{array}
			\right)$      &      $\left(
			\begin{array}{cc}
				-1.374&  1.594\\ -1.114&1.536\\  0.797&  0.246\\
			\end{array}
			\right)$  \\  \hline
			\rowcolor[HTML]{FFFFFF} 
			\multicolumn{1}{|c|}{\cellcolor[HTML]{FFFFFF}
				${y^{q}_c}$   }                                                                & $\left(
			\begin{array}{cc}
				0.233 &0.251 \\0.279 & 0.376 \\ 1.569 &1.849\\
			\end{array}
			\right)  $   &      $\left(
			\begin{array}{cc}
				0.913 &0.452 \\0.555 & 0.265 \\ 0.534 &0.229\\
			\end{array}
			\right)  $ \\
			\hline
			\rowcolor[HTML]{FFFFFF} 
			\multicolumn{1}{|c|}{\cellcolor[HTML]{FFFFFF}
				${y^{q}_1}$   }                                                                & $\left(
			\begin{array}{c}
				1.21 \times e^{0.060 i \vphantom{10^10}} \\
				1.55  \\
			\end{array}
			\right)  $   &     $\left(
			\begin{array}{c}
				0.300 \times e^{0.060 i \vphantom{10^10}} \\
				-0.200  \\
			\end{array}
			\right)  $ \\
			\hline
			\rowcolor[HTML]{FFFFFF} 
			\multicolumn{1}{|c|}{\cellcolor[HTML]{FFFFFF}
				${y^{q}_3}$   }                                                                & $\left(
			\begin{array}{c}
				0.740 \vphantom{10^10}\\
				1.320  \\
			\end{array}
			\right)  $   &       $\left(
			\begin{array}{c}
				1.200 \vphantom{10^10}  \\
				0.800  \\
			\end{array}
			\right)  $ \\
			\hline\hline
			\rowcolor[HTML]{FFFFFF} 
			\multicolumn{1}{|c|}{\cellcolor[HTML]{FFFFFF}
				$y^{\ell}_a$                 }                                                  &  $\left(
			\begin{array}{cc}
				0.267 & 0.272 \vphantom{10^10}\\0.752 & 0.730 \\ 0.583 &0.580\\
			\end{array}
			\right)  $  &  $\left(
			\begin{array}{cc}
				1.217 & 1.143 \vphantom{10^10} \\0.555 & 0.519 \\ 0.552 &0.512\\
			\end{array}
			\right)  $     \\  \hline
			\rowcolor[HTML]{FFFFFF} 
			\multicolumn{1}{|c|}{\cellcolor[HTML]{FFFFFF}
				$y^{\ell}_c$                 }                                                  &  $\left(
			\begin{array}{cc}
				0.580 & 0.371 \vphantom{10^10} \\0.860 & 0.638 \\ 0.587 &0.430\\
			\end{array}
			\right)  $   &     $\left(
			\begin{array}{cc}
				0.758 & 0.288 \vphantom{10^10} \\0.754 & 0.246 \\ 0.277 &0.123\\
			\end{array}
			\right)   $   \\ 
			\hline
			\rowcolor[HTML]{FFFFFF} 
			\multicolumn{1}{|c|}{\cellcolor[HTML]{FFFFFF}
				${y^{\ell}_1}$   }                                                                & $\left(
			\begin{array}{c}
				0.650 \vphantom{10^10} \\
				0.920 \\
			\end{array}
			\right)  $   &     $ \left(
			\begin{array}{c}
				0.600 \vphantom{10^10} \\
				0.700  \\
			\end{array}
			\right)  $\\
			\hline
			\rowcolor[HTML]{FFFFFF} 
			\multicolumn{1}{|c|}{\cellcolor[HTML]{FFFFFF}
				${y^{\ell}_3}$   }                                                                & $\left(
			\begin{array}{c}
				1.680 \vphantom{10^10}\\
				0.740  \\
			\end{array}
			\right)  $   &     $\left(
			\begin{array}{c}
				0.500 \vphantom{10^10} \\
				0.100  \\
			\end{array}
			\right)  $ \\
			\hline
			\hline
		\end{tabular}
	}
	\caption{Yukawa coupling benchmark points for our analysis.}
	\label{Tab:yukawabenchmarks}
\end{table}

\begin{table}[htb!]
	\centering 
	\resizebox{0.55\textwidth}{!}{%
	\begin{tabular}{|cccc|}
		\hline \hline
		\multicolumn{4}{|c|}{\bf{Quark Sector}} \\ \hline
		\multicolumn{1}{|c|}{} & \multicolumn{1}{c|}{} & \multicolumn{2}{c|}{\bf{Model Prediction}} \\ \cline{3-4}
		\multicolumn{1}{|c|}{\multirow{-2}{*}{\begin{tabular}[c]{@{}c@{}}\bf{Observable}\\ \bf{(Masses in GeV)}\end{tabular}}} & \multicolumn{1}{c|}{\multirow{-2}{*}{\bf{Exp. Range}}} & \multicolumn{1}{c|}{\bf{BP1}} & \bf{BP2} \\ \hline \hline
		\rowcolor[HTML]{FFFFFF} 
		\multicolumn{1}{|c|}{\cellcolor[HTML]{FFFFFF}$m_u / 10^{-3}$} & \multicolumn{1}{c|}{\cellcolor[HTML]{FFFFFF}$1.38 \rightarrow 3.63$} & \multicolumn{1}{c|}{\cellcolor[HTML]{FFFFFF}$2.17$} & $2.05$ \\ \cline{1-4}
		\rowcolor[HTML]{FFFFFF} 
		\multicolumn{1}{|c|}{\cellcolor[HTML]{FFFFFF}$m_c$} & \multicolumn{1}{c|}{\cellcolor[HTML]{FFFFFF}$1.21 \rightarrow 1.33$} & \multicolumn{1}{c|}{\cellcolor[HTML]{FFFFFF}$1.27$} & $1.23$ \\ \cline{1-4}
		\rowcolor[HTML]{FFFFFF} 
		\multicolumn{1}{|c|}{\cellcolor[HTML]{FFFFFF}$m_t$} & \multicolumn{1}{c|}{\cellcolor[HTML]{FFFFFF}$171.7 \rightarrow 174.1$} & \multicolumn{1}{c|}{\cellcolor[HTML]{FFFFFF}$172.9$} & $172.9$ \\ \hline
		\rowcolor[HTML]{FFFFFF} 
		\multicolumn{1}{|c|}{\cellcolor[HTML]{FFFFFF}$m_d/ 10^{-3}$} & \multicolumn{1}{c|}{\cellcolor[HTML]{FFFFFF}$4.16 \rightarrow 6.11$} & \multicolumn{1}{c|}{\cellcolor[HTML]{FFFFFF}$4.65$} & $4.73$ \\ \cline{1-4}
		\rowcolor[HTML]{FFFFFF} 
		\multicolumn{1}{|c|}{\cellcolor[HTML]{FFFFFF}$m_s$} & \multicolumn{1}{c|}{\cellcolor[HTML]{FFFFFF}$0.078 \rightarrow 0.126$} & \multicolumn{1}{c|}{\cellcolor[HTML]{FFFFFF}$0.094$} & $0.120$ \\ \cline{1-4}
		\rowcolor[HTML]{FFFFFF} 
		\multicolumn{1}{|c|}{\cellcolor[HTML]{FFFFFF}$m_b$} & \multicolumn{1}{c|}{\cellcolor[HTML]{FFFFFF}$4.12 \rightarrow 4.27$} & \multicolumn{1}{c|}{\cellcolor[HTML]{FFFFFF}$4.18$} & $4.18$ \\ \hline
		\rowcolor[HTML]{FFFFFF} 
		\multicolumn{1}{|c|}{\cellcolor[HTML]{FFFFFF}$|V_{\mathrm{ud}}|$} & \multicolumn{1}{c|}{\cellcolor[HTML]{FFFFFF}$0.973 \rightarrow 0.974$} & \multicolumn{1}{c|}{\cellcolor[HTML]{FFFFFF}$0.974$} & $0.974$ \\ \cline{1-4}
		\rowcolor[HTML]{FFFFFF} 
		\multicolumn{1}{|c|}{\cellcolor[HTML]{FFFFFF}$|V_{\mathrm{us}}|$} & \multicolumn{1}{c|}{\cellcolor[HTML]{FFFFFF}$0.222 \rightarrow 0.227$} & \multicolumn{1}{c|}{\cellcolor[HTML]{FFFFFF}$0.227$} & $0.226$ \\ \cline{1-4}
		\rowcolor[HTML]{FFFFFF} 
		\multicolumn{1}{|c|}{\cellcolor[HTML]{FFFFFF}$|V_{\mathrm{ub}}| /10^{-4}$} & \multicolumn{1}{c|}{\cellcolor[HTML]{FFFFFF}$31.0 \rightarrow 45.4$} & \multicolumn{1}{c|}{\cellcolor[HTML]{FFFFFF}$38.7$} & $39.0$ \\ \hline
		\rowcolor[HTML]{FFFFFF} 
		\multicolumn{1}{|c|}{\cellcolor[HTML]{FFFFFF}$|V_{\mathrm{cd}}|$} & \multicolumn{1}{c|}{\cellcolor[HTML]{FFFFFF}$0.209 \rightarrow 0.233$} & \multicolumn{1}{c|}{\cellcolor[HTML]{FFFFFF}$0.226$} & $0.226$ \\ \cline{1-4}
		\rowcolor[HTML]{FFFFFF} 
		\multicolumn{1}{|c|}{\cellcolor[HTML]{FFFFFF}$|V_{\mathrm{cs}}|$} & \multicolumn{1}{c|}{\cellcolor[HTML]{FFFFFF}$0.954 \rightarrow 1.020$} & \multicolumn{1}{c|}{\cellcolor[HTML]{FFFFFF}$0.973$} & $0.973$ \\ \cline{1-4}
		\rowcolor[HTML]{FFFFFF} 
		\multicolumn{1}{|c|}{\cellcolor[HTML]{FFFFFF}$|V_{\mathrm{cb}}| / 10^{-3}$} & \multicolumn{1}{c|}{\cellcolor[HTML]{FFFFFF}$36.8 \rightarrow 45.2$} & \multicolumn{1}{c|}{\cellcolor[HTML]{FFFFFF}$40.0$} & $40.3$ \\ \hline
		\rowcolor[HTML]{FFFFFF} 
		\multicolumn{1}{|c|}{\cellcolor[HTML]{FFFFFF}$|V_{\mathrm{td}}| / 10^{-4}$} & \multicolumn{1}{c|}{\cellcolor[HTML]{FFFFFF}$71.0 \rightarrow 89.0$} & \multicolumn{1}{c|}{\cellcolor[HTML]{FFFFFF}$80.6$} & $80.4$ \\ \cline{1-4}
		\rowcolor[HTML]{FFFFFF} 
		\multicolumn{1}{|c|}{\cellcolor[HTML]{FFFFFF}$|V_{\mathrm{ts}}| /10^{-3}$} & \multicolumn{1}{c|}{\cellcolor[HTML]{FFFFFF}$35.5 \rightarrow 42.1$} & \multicolumn{1}{c|}{\cellcolor[HTML]{FFFFFF}$39.4$} & $39.6$ \\ \cline{1-4}
		\rowcolor[HTML]{FFFFFF} 
		\multicolumn{1}{|c|}{\cellcolor[HTML]{FFFFFF}$|V_{\mathrm{tb}}|$} & \multicolumn{1}{c|}{\cellcolor[HTML]{FFFFFF}$0.923 \rightarrow 1.103$} & \multicolumn{1}{c|}{\cellcolor[HTML]{FFFFFF}$0.999$} & $0.999$ \\ \hline
		\rowcolor[HTML]{FFFFFF} 
		\multicolumn{1}{|c|}{\cellcolor[HTML]{FFFFFF}$\mathcal{J}/ 10^{-5}$} & \multicolumn{1}{c|}{\cellcolor[HTML]{FFFFFF}$2.73 \rightarrow 3.45$} & \multicolumn{1}{c|}{\cellcolor[HTML]{FFFFFF}$3.06$} & $3.05$ \\ \hline
		\multicolumn{4}{|c|}{\bf{Charged Lepton Sector}} \\ \hline
		\rowcolor[HTML]{FFFFFF} 
		\multicolumn{1}{|c|}{\cellcolor[HTML]{FFFFFF}$m_e /10^{-3}$} & \multicolumn{1}{c|}{\cellcolor[HTML]{FFFFFF}$0.485 \rightarrow 0.537$} & \multicolumn{1}{c|}{\cellcolor[HTML]{FFFFFF}$0.511$} & $0.512$ \\ \cline{1-4}
		\rowcolor[HTML]{FFFFFF} 
		\multicolumn{1}{|c|}{\cellcolor[HTML]{FFFFFF}$m_\mu$} & \multicolumn{1}{c|}{\cellcolor[HTML]{FFFFFF}$0.100 \rightarrow 0.111$} & \multicolumn{1}{c|}{\cellcolor[HTML]{FFFFFF}$0.106$} & $0.106$ \\ \cline{1-4}
		\rowcolor[HTML]{FFFFFF} 
		\multicolumn{1}{|c|}{\cellcolor[HTML]{FFFFFF}$m_\tau$} & \multicolumn{1}{c|}{\cellcolor[HTML]{FFFFFF}$1.688 \rightarrow 1.866$} & \multicolumn{1}{c|}{\cellcolor[HTML]{FFFFFF}$1.777$} & $1.788$ \\ \cline{1-4}
	\end{tabular}
	}
	\caption{Observables for the given benchmark points. We display the $3\sigma$ experimental ranges from \cite{Zyla:2020zbs, Esteban:2020cvm} with the exception of charged lepton masses where we indicate the $\pm 5\%$ range. }
	\label{Tab.: Masses and Mixings with BPs}
\end{table}
In order to demonstrate the viability of this model, we start with two benchmark scenarios that can reproduce the observed mass hierarchies and mixings in the quark sector as well as the charged lepton masses. In both scenarios, we assume $\alpha=1/2$ and $g_X = 1$. For the first benchmark point (BP) we consider the breaking scales $v_R=20 $ TeV and $v_\eta = 30$ TeV. Note that this implies the new gauge bosons masses  $M_{W_R} = 6.53~\mathrm{TeV}$ and $M_{Z'}= 7.82 ~\mathrm{TeV}$. Vector-like quark masses are given by $M_{T1}=9$ TeV, $M_{T2}=14$ TeV, $M_{B1}=38$ TeV and $M_{B2}=42$ TeV, whereas in the lepton sector we consider $M_{N1}=60$ TeV, $M_{N2}=65$ TeV, $M_{E1}=44$ TeV and $M_{E2}=51$ TeV.
For the second benchmark point, we assume larger symmetry-breaking scales $v_R=80 $ TeV and $v_\eta = 120$ TeV. The gauge boson masses are therefore given by $M_{W_R} = 26.13~\mathrm{TeV}$ and $M_{Z'}= 31.26 ~\mathrm{TeV}$, respectively.  The vector-like fermion masses are also slightly larger with  $M_{T1}=19$ TeV, $M_{T2}=21$ TeV, $M_{B1}=47$ TeV and  $M_{B2}=49$ TeV in the quark sector, and $M_{N1}=82$ TeV, $M_{N2}=85$ TeV, $M_{E1}=61$ TeV and $M_{E2}=64$ TeV in the lepton sector. Table~\ref{Tab:yukawabenchmarks} presents the two sets of Yukawa couplings used in our numeric calculation. Note that the given parameters suffice to describe Dirac masses for all fermions and LNV contributions in the neutrino sector will be considered later in this section. 

In Table~\ref{Tab.: Masses and Mixings with BPs}, we demonstrate that the two given BP's reproduce quark masses and mixings, as well as charged lepton masses, consistent with current experimental bounds. Observables in the quark sector agree with the $3\sigma$ experimental ranges, and charged lepton masses match within $\pm 5\%$ with the current best-fit values.  We highlight that this is achieved with Yukawa couplings of order one, and the hierarchy within the generations arises mainly from loop factors. Apart from that, differing mass scales for the top, bottom, and tau arise due to a mild hierarchy of vector-like masses $M_T < M_B < M_E$. Moreover, light Dirac neutrino masses are at the order of charged lepton masses, since $M_N \simeq M_E$ and Yukawa couplings are $\mathcal{O}(1)$ in the considered BP's.

In Section~\ref{Sec.: Neutrino Masses}, we discuss possible LNV contributions, and one has to show that they can lead to sub-eV active neutrino masses. The behavior of the full $12 \times 12$ mass matrix given in Eq.~(\ref{Eq.: Neutrino mass Complete tree level}) is complicated and we therefore try to answer the question for the third-generation case (where $i=3$ in  Eq.~(\ref{Eq.: Neutrino mass Complete tree level})). While we retain the lepton number conserving couplings and masses from BP1 and BP2, we examine which LNV masses lead to a light neutrino with  $m_\nu < 0.3$ eV. For that we assume $M'_{L,11}=M'_{L,22} \equiv M'_{L} $,  $M'_{L,12}=0 $ (and similarly for $L \leftrightarrow R$) and randomly scatter $M'_L$ and $M'_R$ in an interval $\{10^3, ~10^{14}\}$ GeV. The parameter points that yield a sub-eV neutrino are displayed in Fig.~\ref{Fig.:neutrino scatter}. We distinguish the two separate scenarios, with vanishing LNV couplings $y' = 0$ (left panel)  and non-zero couplings $y'\neq 0$ randomly chosen in an interval $\{0.1,~1.0\}$ (right panel). Evidently, non-zero $y'$ opens up a huge parameter space compared to the more restrictive assumption $y' = 0$. 

Counting parameters in the LNV sector, it seems possible to realize small neutrino masses for all three active neutrinos. A full study with all generations would then also need to include radiative corrections that contribute to both LNV and lepton number conserving terms. It is quite likely that subtle dependencies within the $12 \times 12$ matrix need to be tuned to achieve active neutrino masses in agreement with current oscillation data. Our preliminary numerical estimates suggest that the existence of four light neutrinos is possible, with one potentially being a sterile neutrino at the eV scale or higher, which could have important phenomenological implications. Therefore, we plan to undertake a comprehensive parameter scan and an in-depth study of light sterile neutrino phenomenology in future work.
\section{Conclusion}
In this paper, we have presented a scheme for explaining the mass hierarchy of standard model charged fermions by extending the SM gauge group to the left-right symmetric one and using the universal seesaw mechanism to give the masses to the third generation fermions at the tree-level and, which at one and two loop levels give masses to second and first generation fermions respectively. The loop suppression essentially explains the mass hierarchy as well as the inversion of up and down quark masses of the first generation. Our work does not use any new scalars, which would have introduced more parameters into the model, and instead, it relies only on the gauge interactions of the left-right symmetric theory and the interactions of the  Higgs field that are needed to break the gauge symmetry. We also give a fit to quark mixing angles and CP violation for two benchmark choices of the model parameters.
\section*{Acknowledgement}
We thank K.S. Babu and Anil Thapa for useful discussions.
\appendix
\renewcommand\thesection{\Alph{section}.}
\renewcommand\thesubsection{\thesection\arabic{subsection}}
\section*{Appendix}
\addcontentsline{toc}{section}{Appendix}
\section*{A. Gauge boson and scalar masses}
\label{App.: Gauge Boson and Scalar Masses}
The electrically neutral components of the scalars $\chi_L= (\chi_L^+, \chi_L^0)^T$, $\chi_R =(\chi_R^+, \chi_R^0)^T$ and $\eta$ are given by:
\begin{equation}
	\chi_L^0  = \dfrac{1}{\sqrt{2}}\left(
		v_{L}+ \sigma_{L}+ i \rho_{ L}
	\right) \, ,~
	\chi_R^0  = \dfrac{1}{\sqrt{2}}\left(
		v_{R}+ \sigma_R + i \rho_{R}
	\right) \, ,~
	 \eta  = \dfrac{1}{ \sqrt{2}}\left( v_\eta + \sigma_\eta + i \rho_{\eta}\right) \, .
\end{equation}
where $v_L$, $v_R$ and $v_\eta$ denote the vacuum expectation values, respectively. We consider all the couplings in the scalar potential of Eq.(\ref{Eq.:Scalar Potential}) to be real. The  mass matrix for the real scalar fields is then given by
\begin{equation}
    \mathcal{L} \supset \dfrac{1}{2}
    \left(
    \begin{array}{ccc}
         \sigma_L &
         \sigma_R&
         \sigma_\eta \\
    \end{array}
    \right) 
    \mathcal{M}^2_\sigma
    \left(
    \begin{array}{c}
         \sigma_L \\
         \sigma_R\\
         \sigma_\eta \\
    \end{array}
    \right) ~,
\end{equation}
where we defined
\begin{equation}
    \mathcal{M}^2_\sigma = 
    \left(
    \begin{array}{ccc}
        2\lambda_1 v_L^2  & \lambda_2 v_L v_R  &  \lambda_4 v_L v_\eta \\
         \lambda_2 v_L v_R & 2\lambda_1 v_R^2  &  \lambda_4 v_R v_\eta \\
          \lambda_4 v_L v_\eta&  \lambda_4 v_R v_\eta & 2\lambda_3 v_\eta^2 \\
    \end{array}
    \right)~.
\end{equation}
The fields $\rho_{L}$, $\rho_{R}$ and $\rho_{\eta}$ are the massless Goldstone modes which become the longitudinal components of the $Z_L, Z_R$, and $Z_X$ bosons after spontaneous symmetry breaking.
We consider the limit of vanishing coupling $\lambda_4 \ll 1$, such that $\eta$ decouples and  has a mass
$$
M_\eta^2 = 2 \lambda_3 v_\eta^2
$$
Diagonalizing the $\sigma_L-\sigma_R$ subspace of the neutral scalar mass matrix yields the two mass eigenstates
\begin{equation}
\left(
\begin{array}{c}
     h  \\
     H 
\end{array}
\right) = 
\left(
\begin{array}{cc}
     c_\xi & s_\xi \\
     -s_\xi & c_\xi \\
\end{array}
\right) 
\left(
\begin{array}{c}
     \sigma_L  \\
     \sigma_R 
\end{array}
\right)
\end{equation}
where $s_\xi \equiv \sin(\xi)$ with 
$$
\tan(2 \xi) = \dfrac{\lambda_2 v_L v_R}{\lambda_1(v_R^2-v_L^2)} ~.
$$
In the limit $v_L \ll v_R$, the mass eigenvalues are approximately given by
\begin{align}
    M_h^2 &\simeq \left(2 \lambda_1 - \dfrac{\lambda_2^2}{2 \lambda_1} \right) v_L^2 ~ ,\\
     M_H^2 &\simeq 2 \lambda_1 v_R^2 ~ .
\end{align}
If we fix the Higgs mass $M_h$ to its SM value, we can constrain $\lambda_1$ as:
\begin{equation}
    \lambda_1 = \dfrac{1}{8}\left( \dfrac{2 M_{h}^2}{v_L^2} + \sqrt{\left(\dfrac{2 M_{h}^2}{v_L^2}\right)^2+16 \lambda_2^2}\right)~.
\end{equation}
Out of the eight gauge bosons that arise from the $SU(2)\times SU(2) \times U(1) \times U(1)$ gauge symmetry, seven become massive via spontaneous symmetry breaking. In our convention, the gauge bosons $W^i_{L \mu}$ $\left(W^i_{R\mu} \right)$, $i=1,2,3$ are assiciated with $SU(2)_L$ $\left(SU(2)_R \right)$, while $B_\mu$ and $X_\mu$ belong to $U(1)_X$ and $U(1)_{X'}$, respectively. The gauge boson mass terms arise from the kinetic terms
\begin{equation}
\label{Eq.: Kinetic Lagrangian scalars}
	\mathcal{L} \supset \left(D_\mu \chi_L \right)^\dagger \left(D^\mu \chi_L \right) + \left(D_\mu \chi_R \right)^\dagger \left(D^\mu \chi_R \right) + \left(D_\mu \eta \right)^\dagger \left(D^\mu \eta \right)  ~,
\end{equation}
after the scalars acquire non-zero VEVs
\begin{equation}
	\langle \chi_L\rangle = \dfrac{1}{\sqrt{2}}\left(
	\begin{array}{c}
		0 \\
		v_L \\
	\end{array}
	\right) ~, \indent 
	\langle \chi_R\rangle = \dfrac{1}{\sqrt{2}}\left(
	\begin{array}{c}
		0 \\
		v_R \\
	\end{array}
	\right) ~,  \indent 
	\langle \eta\rangle = \dfrac{1}{\sqrt{2}}v_\eta ~.
\end{equation}
The charged gauge bosons $W^\pm_L = \left(W^2_L \mp i W^1_L\right)/\sqrt{2} $ and $W^\pm_R = \left(W^2_R \mp i W^1_R\right)/\sqrt{2}$ obtain a mass
\begin{equation}
	M^2_{W_L} = \dfrac{g_L^2 v_L^2}{4} ~, \indent 	M^2_{W_R} = \dfrac{g_R^2 v_L^2}{4} ~.
\end{equation}
Here, the couplings $g_L$ and $g_R$ correspond to the gauge groups $SU(2)_L$ and $SU(2)_R$. The gauge couplings for the groups $U(1)_X$ and  $U(1)_{X'}$ are labeled as $g'$ and $g_X$ in the following.

The squared mass matrix for the neutral gauge bosons in the basis $(W^3_L, W^3_R, B, X)$ is given by
\begin{equation}
	\mathcal{M}^2 = \dfrac{1}{4} \left(
	\begin{array}{cccc}
		g_L^2 v_L^2  &   0   &   -g_L g' v_L^2  & 0\\
		0            &   g_R^2 v_R^2  &  -g_R g' v_R^2 &0 \\
		-g_L g' v_L^2    &  -g_R g' v_R^2   & g'^2 (v_R^2 + v_L^2)  &0\\
		0&0&0&  g_X^2 v_\eta^2\\
	\end{array}
	\right) ~.
\end{equation}
Evidently, the $X $ gauge boson decouples from the spectrum and has a mass
$$
M_{X}^2 = \dfrac{g_x^2 v_\eta^2}{4} \, .
$$
Since we assume the  Lagrangian to be parity symmetric, left -and right-handed gauge couplings are similar and we write $g \equiv g_L = g_R$ in the following. By defining the rotation angle $s_w \equiv \sin(\theta_w)= e/g$ with $1/e^2 = 1/g_R^2+1/g_L^2+ 1/g'^2$, it can be shown that a transformation to the basis $(A, Z_L, Z_R)$ defined by 
\begin{equation}
	\left(
	\begin{array}{c}
		A \\
		Z_L \\
		Z_R \\
	\end{array}
	\right)
	=  \left(
	\begin{array}{ccc}
		s_w  &   s_w   &  \sqrt{c_w^2 - s_w^2} \\
		c_w            &   -s_w^2 /c_w & -s_w \sqrt{c_w^2-s_w^2}/c_w \\
		0    &  \sqrt{c_w^2-s_w^2}/c_w   & -s_w /c_w  \\
	\end{array}
	\right)
	\left(
	\begin{array}{c}
		W_L^3 \\
		W_R^3 \\
		B \\
	\end{array}
	\right)
	~.
\end{equation}
yields one zero eigenvalue in the mass matrix. This massless gauge boson $A$ can be identified with the photon. The two by two mass matrix which  spans the subspace $(Z_L, Z_R)$ is then given by
\begin{equation}
	\begin{split}
		M_{LL}^2 &= \dfrac{e^2 v_L^2}{4 s_w^2 c_w^2} ~,\\
		M_{RL}^2 &=   \dfrac{e^2 v_L^2}{4  c_w^2 \sqrt{c_w^2-s_w^2}} ~, \\
		M_{RR}^2 &=  \dfrac{e^2}{4 \left(c_w^2-s_w^2\right)}\left(\dfrac{c_w^2 v_R^2}{s_w^2}+ \dfrac{s_w^2 v_L^2}{c_w^2}\right) ~. \\
	\end{split}
\end{equation}
Finally, rotating this to the mass eigenstates $(Z, Z')$ by the transformation
\begin{equation}
	\left(
	\begin{array}{c}
		Z \\
		Z' \\
	\end{array}
	\right)
	=  \left(
	\begin{array}{cc}
		c_\zeta  &   s_\zeta    \\
		-s_\zeta           &   c_\zeta  \\
	\end{array}
	\right)
	\left(
	\begin{array}{c}
		Z_L \\
		Z_R \\
	\end{array}
	\right)
	~,
\end{equation}
with a mixing angle 
\begin{equation}
\label{Eq.: ZL-ZR Mixing Angle}
	\tan(2\zeta) = \dfrac{2 M_{LR}^2}{M_{LL}^2- M_{RR}^2} ~.
\end{equation}
Following from this, the two mass eigenvalues are
\begin{equation}
	M_{Z, Z'}^2 = \dfrac{1}{2} \left(M_{LL}^2+ M_{RR}^2 \mp (M_{LL}^2- M_{RR}^2)\sqrt{1+\tan^2(2\zeta)}\right) ~,
\end{equation}
where we identify the lighter state with the SM $Z$ boson.
\section*{B. Fermion interaction}
Diagonalizing the mass matrix, including one-loop and two-loop radiative corrections by a bi-unitary transformation, yields
\begin{equation}
\begin{split}
    V_L^u \mathcal{M}_u^{(2)} (V_R^u)^{\dagger} &=  \mathcal{M}^{\mathrm{diag}}_u \equiv \mathrm{diag}(m_u, m_c, m_t , m_{U_1}, m_{U_2},  m_{U_3}) \, ,\\
    V_L^d \mathcal{M}_d^{(0)} (V_R^d)^{\dagger} &=  \mathcal{M}^{\mathrm{diag}}_d \equiv \mathrm{diag}(m_d,m_s, m_b , m_{D_1}, m_{D_2},  m_{D_3}) \, ,
    \end{split}
\end{equation}
where the mass eigenstates are given by
\begin{equation}
\label{Eq.: quark mass eigensate}
    \begin{split}
        \hat{\mathbf{u}}_{L/ R} = V^u_{L/R} \mathbf{u}_{L/R} \, ,\\
         \hat{\mathbf{d}}_{L/ R} = V^d_{L/R} \mathbf{d}_{L/R} \, .
    \end{split}
\end{equation}
In this new basis, the $W_L$ boson couples to quarks via 
\begin{equation}
    \mathcal{L} \supset \dfrac{g}{\sqrt{2}}W^+_{L \mu} \left[\overline{\hat{\mathbf{u}}} \gamma^\mu \hat{g}_L^q(W_L) P_L \hat{\mathbf{d}}  \right] + h.c. \, , 
\end{equation}
with
\begin{equation}
    \begin{split}
        \hat{g}_L^q(W_L) &= V_L^u \,  g_L^q(W_L) \, ( V_L^d)^\dagger \, ,
    \end{split}
\end{equation}
and 
\begin{equation}
    \begin{split}
         g_L^q(W_L) &=  \mathrm{diag}(1,1,1,0, 0, 0)\, .
    \end{split}
\end{equation}
Hence, the left-handed CKM mixing matrix is given by
\begin{equation}
    U_{L} \equiv  \hat{g}_L^q(W_L)  \, .
\end{equation}
Additionally, we define the right-handed analog of the CKM matrix as
\begin{equation}
    U_R \equiv  \hat{g}_R^q(W_R) = V^u_R g_R^q(W_R) (V_R^d)^\dagger \, ,
\end{equation}
with $g_R^q(W_R) = \mathrm{diag}(1,1,1,0,0,0)$.

The couplings of the $Z_L$ and $Z_R$ to up-type quarks are given by
\begin{equation}
    \begin{split}
        \mathcal{L}  &\supset  Z_{L\mu} \overline{\mathbf{u}} \gamma^\mu \left[  g_L^u(Z_L) P_L +   g_R^u(Z_L)  P_R\right]\mathbf{u}\\
        &+  Z_{R\mu} \overline{\mathbf{u}}\gamma^\mu \left[  g_L^u(Z_R) P_L +   g_R^u(Z_R)  P_R\right]\mathbf{u}\, ,
    \end{split}
\end{equation}
where we defined the operators 
\begin{equation}
\begin{split}
    g_L^u(Z_L) &=\dfrac{g}{c_w} \left[ T_L^3 - Q s^2_w \right] \, , \indent
    g_R^u(Z_L) =\dfrac{g}{c_w}\left[ - Q s^2_w \right] \, ,\\
    g_L^u(Z_R) &=\dfrac{g c_w}{\sqrt{c^2_w-s_w^2}} \left[  t^2_w \left(T_L^3 -Q \right) \right] \, ,\indent
    g_R^u(Z_R) =\dfrac{g c_w}{\sqrt{c^2_w-s_w^2}}\left[T_R^3 - t^2_w  Q \right] \, ,
    \end{split}
\end{equation}
acting diagonally on each generation according to their  electric charge $Q$ and third component of $SU(2)_L$ ($SU(2)_R$) isospin $T_L^3$ ($T_R^3$). Note that we use the abbreviation $t_w \equiv s_w/c_w$. Rotating this to the gauge boson mass eigenbasis $(Z, ~Z')$ yields
\begin{equation}
    \begin{split}
        \mathcal{L}  &\supset g Z_{\mu} \overline{\mathbf{u}} \gamma^\mu \left[  c_\zeta\{ g_L^u(Z_L) P_L +   g_R^u(Z_L)  P_R\} + s_\zeta\{ g_L^u(Z_R) P_L +    g_R^u(Z_R)  P_R\}\right]\mathbf{u}\\
        &+ g Z'_{\mu} \overline{\mathbf{u}}\gamma^\mu \left[  -s_\zeta\{ g_L^u(Z_L) P_L +   g_R^u(Z_L)  P_R\}+ c_\zeta\{ g_L^u(Z_R) P_L +    g_R^u(Z_R)  P_R\}\right]\mathbf{u} ~.
\end{split}
\end{equation}
Finally, going also to the fermion mass basis, the Lagrangian becomes
\begin{equation}
    \begin{split}
        \mathcal{L}  &\supset g Z_{\mu} \overline{\mathbf{\hat{u}}} \gamma^\mu \left[  \hat{C}_L(Z) P_L + \hat{C}_R(Z) P_R\right]\mathbf{\hat{u}}\\
        &+ g Z'_{\mu} \overline{\mathbf{\hat{u}}}\gamma^\mu \left[  \hat{C}_L(Z') P_L + \hat{C}_R(Z')P_R\right]\mathbf{\hat{u}} 
\end{split}
\end{equation}
where we defined
\begin{equation}
    \begin{split}
        C_L(Z) &\equiv  c_\zeta g_L^u(Z_L) + s_\zeta g_L^u(Z_R)    ~,\\
          C_R(Z)  &\equiv     c_\zeta g_R^u(Z_L) +  s_\zeta g_R^u(Z_R)   ~,\\
          C_L(Z')  &\equiv  -s_\zeta g_L^u(Z_L) + c_\zeta g_L^u(Z_R)    ~,\\
         C_R(Z') &\equiv     -s_\zeta g_R^u(Z_L) +  c_\zeta g_R^u(Z_R)  ~, 
    \end{split}
\end{equation}
and
\begin{equation}
    \begin{split}
        \hat{C}_L(Z) &\equiv V_L C_L(Z) V_L^\dagger ~,\\
        \hat{C}_R(Z) &\equiv V_R C_R(Z) V_R^\dagger ~,\\
        \hat{C}_L(Z') &\equiv V_L C_L(Z') V_L^\dagger ~,\\
        \hat{C}_R(Z') &\equiv V_R C_R(Z') V_R^\dagger ~.
    \end{split}
\end{equation}
Note that the previous couplings are not proportional to the identity matrix since the SM quarks mix with the vector-like quarks, which have different gauge charges.

The interaction of an up-type fermion with the scalars $\sigma_L$, $\sigma_R$ is given by
\begin{equation}
\label{Eq.: Scalar yukawa couplings}
    \mathcal{L} \supset -\dfrac{1}{\sqrt{2}} \sigma_L \overline{\mathbf{u}} (\mathbf{Y_a} P_R+\mathbf{Y_a^\dagger} P_L) \mathbf{u}
    -\dfrac{1}{\sqrt{2}} \sigma_R \overline{\mathbf{u}} (\mathbf{Y_a^\dagger} P_R +\mathbf{Y_a} P_L) \mathbf{u}
    ~,
\end{equation}
where we defined the matrix 
\begin{equation}
    \mathbf{Y_a} \equiv \left(
    \begin{array}{ccc}
       \mathbf{0_{3 \times 3}} &y_a & \mathbf{0_{3 \times 1}} \\
       \mathbf{0_{2 \times 3}} &\mathbf{0_{2 \times 2}} & \mathbf{0_{2 \times 1}} \\
       \mathbf{0_{1 \times 3}}& \mathbf{0_{1 \times 2}}& \mathbf{0_{1 \times 1}}
    \end{array} \right)~.
\end{equation}
If we rotate to the scalar mass eigenbasis, we find
\begin{equation}
    \begin{split}
    \mathcal{L} \supset -\dfrac{1}{\sqrt{2}}  h \overline{\mathbf{u}} (\left(c_\xi\mathbf{Y_a} +s_\xi \mathbf{Y_a^\dagger}\right) P_R +\left(c_\xi\mathbf{Y_a^\dagger} +s_\xi \mathbf{Y_a}\right) P_L) \mathbf{u}\\
    -\dfrac{1}{\sqrt{2}} H \overline{\mathbf{u}} (\left(c_\xi \mathbf{Y_a^\dagger}  -s_\xi\mathbf{Y_a}\right) P_R +\left(c_\xi \mathbf{Y_a} -s_\xi\mathbf{Y_a^\dagger} \right)P_L) \mathbf{u}
    ~,
    \end{split}
\end{equation}
Going to the fermion mass basis yields
\begin{equation}
    \begin{split}
    \mathcal{L} \supset -\dfrac{1}{\sqrt{2}}  h \overline{\mathbf{\hat{u}}} (V_L\left(c_\xi\mathbf{Y_a} +s_\xi \mathbf{Y_a^\dagger}\right) V_R^\dagger P_R +V_R\left(c_\xi\mathbf{Y_a^\dagger} +s_\xi \mathbf{Y_a}\right)V_L^\dagger P_L) \mathbf{\hat{u}}\\
    -\dfrac{1}{\sqrt{2}} H \overline{\mathbf{\hat{u}}} (V_L\left(c_\xi \mathbf{Y_a^\dagger}  -s_\xi\mathbf{Y_a}\right)V_R^\dagger P_R +V_R\left(c_\xi \mathbf{Y_a} -s_\xi\mathbf{Y_a^\dagger} \right)V_L^\dagger P_L) \mathbf{\hat{u}}
    ~,
    \end{split}
\end{equation}
which can be written in simplified form
\begin{equation}
    \begin{split}
    \mathcal{L} \supset -&\dfrac{1}{\sqrt{2}}  h \overline{\mathbf{\hat{u}}} (\mathbf{\hat{C}(h)} P_R +\mathbf{\hat{C}^\dagger(h)} P_L) \mathbf{\hat{u}}\\
    -&\dfrac{1}{\sqrt{2}} H \overline{\mathbf{\hat{u}}} (\mathbf{\hat{C}(H) }P_R +\mathbf{\hat{C}^\dagger(H)} P_L) \mathbf{\hat{u}}
    ~,
    \end{split}
\end{equation}
with 
\begin{equation}
    \begin{split}
        \mathbf{C(h)} &= \left[c_\xi \mathbf{Y_a} + s_\xi \mathbf{Y_a^\dagger} \right]  ~,\\
        \mathbf{C(H)}&= \left[-s_\xi \mathbf{Y_a} + c_\xi \mathbf{Y_a^\dagger} \right] ~,
    \end{split}
\end{equation}
\begin{equation}
    \begin{split}
        \mathbf{\hat{C}(h)} &=   V_L \mathbf{C(h)}  V_R^\dagger~,\\
        \mathbf{ \hat{C}(H) }&=   V_L \mathbf{C(H)}  V_R^\dagger~,
    \end{split}
\end{equation}


\section*{C. Loop calculations}\label{App.: Loop calculations}
In the following, we carry out the calculation of the loop diagrams important for our model. For these, we evaluated the Feynman integrals with the help of the Mathematica {\tt Package-X} \cite{Patel:2016fam}. We provide here the formulas for the most frequently used integrals
\begin{equation}
\label{Eq.: Feynman Integral 1}
\begin{split}
    \int \dfrac{d^4 k}{(2 \pi)^4} \dfrac{1}{k^2-m^2} \dfrac{1}{k^2-M^2} &= \dfrac{i}{16 \pi^2} \left(1+ \tilde{\epsilon}- \dfrac{m^2 \log (\frac{m^2}{M^2})}{m^2-M^2} + \log (\mu^2/M^2) \right)~,\\
    \int \dfrac{d^4 k}{(2 \pi)^4} \dfrac{1}{k^2}\dfrac{1}{k^2-m^2} \dfrac{1}{k^2-M^2} &= \dfrac{-i}{16 \pi^2} \dfrac{\log(\frac{m^2}{M^2})}{m^2-M^2}~,
    \\
    \int \dfrac{d^4 k}{(2 \pi)^4} \dfrac{1}{k^2-m^2}\dfrac{1}{k^2-M_1^2} \dfrac{1}{k^2-M_2^2} &= \dfrac{i}{16 \pi^2} \dfrac{1}{M_1^2-M_2^2}\left(-\dfrac{M_1^2 \log(\frac{m^2}{M_1^2})}{m^2-M_1^2}+ \dfrac{M_2^2 \log(\frac{m^2}{M_2^2})}{m^2-M_2^2} \right)~,
    \end{split}
\end{equation}
where we defined the quantity $\tilde{\epsilon} \equiv 1/\epsilon -\gamma_E+\log(4\pi)$.
\subsection*{One-loop corrections}
\begin{figure}[t]
	\centering
	\includegraphics[scale=0.15]{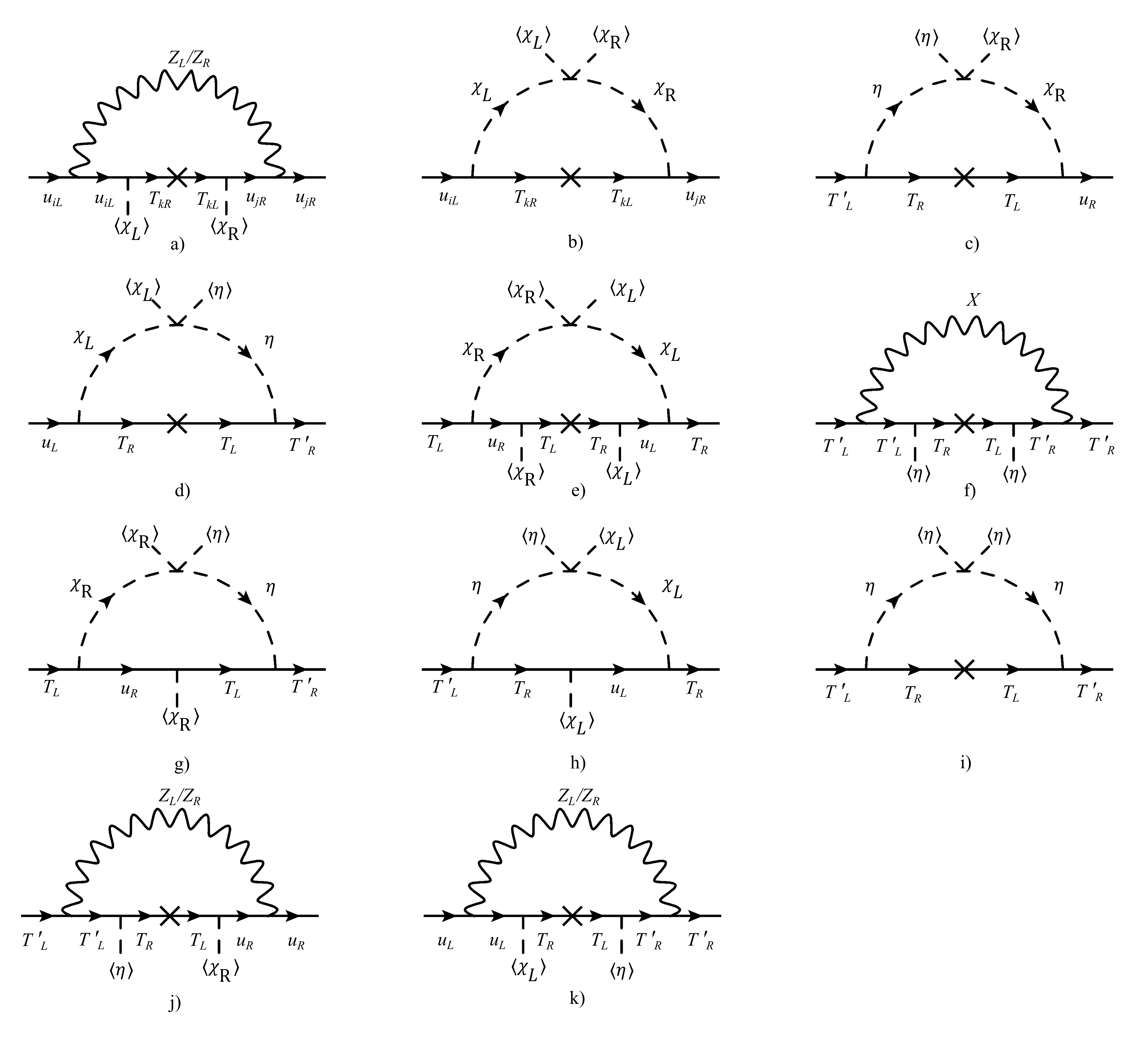}
	\caption{One-loop contributions to up-type quark mass matrix. Out of these diagrams, a) and b) contribute to the zero entry in the upper left corner of the mass matrix.  Contributions from c), d), g), and h) are negligible in the limit $\lambda_4 \ll 1$ where $\eta $ decouples from the scalar sector and hence will not be considered here. Likewise, we neglect the comparatively small contribution from e) to the tree-level vector-like mass $M_T$. Note, however, that the inclusion of those diagrams would not change the matrix rank discussion. The diagrams in j) and k) contribute to the upper right and lower left corners of the mass matrix. Finally, contributions from diagrams f) and i) furnish the lower right zero entry.
 Using the parameterization of Section~\ref{Sec.: eigenstates}, it is easy to verify that at the one-loop level after including all graphs (figures a) through k)), the state $|1\rangle$ remains massless.}
	\label{Fig.: Systematic of one-loop contributions}
\end{figure}

We start with the evaluation of the diagram in Fig.~\ref{Fig.: Systematic of one-loop contributions}~a). The contribution from the exchange of a $Z$ boson mass eigenstate is given by 
\begin{equation}
    \begin{split}
        \delta \mathcal{M}^{\mathrm{(1), Z}}&=  i\int \dfrac{d^4k}{(2 \pi)^4}   C_L(Z)  \gamma^\mu\dfrac{1}{\slashed{k}} \dfrac{y_a v_L} {\sqrt{2}}\dfrac{\left(\slashed{k} + M_T \right)}{k^2 -M_T^2} \dfrac{y^\dagger_a v_R} {\sqrt{2}}\dfrac{1}{\slashed{k}}\\
        &\times \dfrac{1}{(k-p)^2 -M_Z^2}\left[ g_{\mu \nu} -(1-\xi_Z)\dfrac{ (k-p)_\mu (k-p)_\nu}{(k-p)^2-\xi_Z M_Z^2}\right] \gamma^\nu  C_R(Z)  ~.
    \end{split}
\end{equation}
The part of the integral proportional to uneven powers of  $k$ vanishes and we can use the identities  $\gamma^\mu g_{\mu \nu} \gamma^\nu =4 \cdot \identity$, $\slashed{k} \slashed{k}= k^2$ and $\gamma^\mu (k-p)_{\mu} (k-p)_{\nu} \gamma^\nu = (k-p)^2$ to obtain the following result in Landau gauge at zero external momentum $p = 0$:
\begin{equation}
    \begin{split}
        \delta \mathcal{M}^{\mathrm{(1), Z}}&=  i\dfrac{3 C_L(Z) C_R(Z)v_L v_R}{2}\int \dfrac{d^4k}{(2 \pi)^4}     \dfrac{1}{k^2} y_a  \dfrac{ M_T }{k^2 -M_T^2} y^\dagger_a \dfrac{1}{k^2 -M_Z^2}    ~.
    \end{split}
\end{equation}
Using the Feynman integrals from Eq.~(\ref{Eq.: Feynman Integral 1}) we arrive at the final result.
\begin{equation}
    \begin{split}
        \delta \mathcal{M}^{\mathrm{(1), Z}}&=  \dfrac{3 C_L(Z) C_R(Z)v_L v_R}{32 \pi^2} y_a   \dfrac{M_T}{M_T^2-M_Z^2} \log\left(\dfrac{M_T^2}{M_Z^2}\right) y^\dagger_a     ~.
    \end{split}
\end{equation}
Similarly, the contribution by $Z'$ exchange is given by
\begin{equation}
    \begin{split}
        \delta \mathcal{M}^{\mathrm{(1), Z'}}&=  \dfrac{3 C_L(Z') C_R(Z')v_L v_R}{32 \pi^2} y_a   \dfrac{M_T}{M_T^2-M_{Z'}^2} \log\left(\dfrac{M_T^2}{M_{Z'}^2}\right) y^\dagger_a     ~.
    \end{split}
\end{equation}
In principle, one also has to take into account the contributions from the Goldstone bosons $\rho_L$ and $\rho_R$ that are associated with the $Z$ and $Z'$. However, we find that their contribution exactly cancels to zero in the Landau gauge. 

To see this, we quote the relevant part of the Lagrangian that couples the Goldstone bosons to quarks (compare with notation in Eq.~(\ref{Eq.: Scalar yukawa couplings}))
\begin{equation}
    \mathcal{L} \supset -\dfrac{i}{\sqrt{2}} \rho_L \mathbf{\overline{u}} \left(\mathbf{Y_a} P_R + \mathbf{Y_a^\dagger} P_L \right)\mathbf{u}
    -\dfrac{i}{\sqrt{2}} \rho_R \mathbf{\overline{u}} \left(\mathbf{Y_a^\dagger} P_R + \mathbf{Y_a} P_L \right)\mathbf{u}~.
\end{equation}
Bilinears terms of the form $Z^{(')}_\mu \partial^\mu \rho_{L(R)} $ that arise from the kinetic terms in Eq.~(\ref{Eq.: Kinetic Lagrangian scalars}) can be canceled by introducing appropriate gauge fixing terms
\begin{equation}
    \begin{split}
        \mathcal{L}_{\mathrm{GF}} \supset -\dfrac{1}{2\xi_Z} \left(\partial^\mu Z_\mu -i \xi_Z M_Z \hat{\rho}_L \right)^2  -\dfrac{1}{2\xi_{Z'}} \left(\partial^\mu Z'_\mu -i \xi_{Z'} M_{Z'} \hat{\rho}_R \right)^2~,
    \end{split}
\end{equation}
if the following rotation of Goldstone bosons is performed
\begin{equation}
\left(
\begin{array}{c}
     \rho_L  \\
     \rho_R 
\end{array}
\right) = 
\left(
\begin{array}{cc}
     c_G & s_G \\
     -s_G & c_G \\
\end{array}
\right) 
\left(
\begin{array}{c}
     \hat{\rho}_L  \\
     \hat{\rho}_R 
\end{array}
\right)~,
\end{equation}
with $c_G\equiv \cos(\theta_G)$, $s_G\equiv \sin(\theta_G)$ and $\theta_G$ is a mixing angle. The gauge fixing terms are responsible for creating gauge-dependent masses for the Goldstone bosons that are given by
\begin{equation}
    \mathcal{L}_{\mathrm{mass,gold}}=- \dfrac{1}{2} \left( \hat{\rho}_L ~ \hat{\rho}_R\right) \left(
    \begin{array}{cc}
        \xi_Z M_{Z}^2 & 0 \\
        0 & \xi_{Z'} M_{Z'}^2 
    \end{array} \right) \left(
    \begin{array}{cc}
         \hat{\rho}_L \\
        \hat{\rho}_R 
    \end{array} \right)~.
\end{equation}
Now 
defining the couplings on the new basis
\begin{equation}
    \begin{split}
        \mathbf{C(\hat{\rho}_L)} &= \left[c_G \mathbf{Y_a} - s_G \mathbf{Y_a^\dagger} \right]  ~,\\
        \mathbf{C(\hat{\rho}_R)}&= \left[s_G \mathbf{Y_a} + c_G \mathbf{Y_a^\dagger} \right] ~,
    \end{split}
\end{equation}
we can express the Lagrangian as
\begin{equation}
    \mathcal{L} \supset -\dfrac{i}{\sqrt{2}} \hat{\rho}_L \mathbf{\overline{u}} \left(\mathbf{C}(\hat{\rho}_L) P_R + \mathbf{C}(\hat{\rho}_L)^\dagger P_L \right)\mathbf{u}
    -\dfrac{i}{\sqrt{2}} \hat{\rho}_R \mathbf{\overline{u}} \left(\mathbf{C}(\hat{\rho}_R) P_R + \mathbf{C}(\hat{\rho}_R)^\dagger P_L \right)\mathbf{u}
\end{equation}
The Goldstone contribution is hence given by 
\begin{equation}
\label{Eq.: Contribution rho_L }
    \begin{split}
    \delta \mathcal{M}^{\mathrm{1-loop, \hat{\rho}_L}} 
    =& \dfrac{-i}{2} \int \dfrac{d^4k}{(2 \pi)^4} \dfrac{1}{(k-p)^2}  s_G c_G y_a \dfrac{\slashed{k}+ M_T}{k^2 -M_T^2}  y_a^\dagger 
        ~,
    \end{split}
\end{equation}
\begin{equation}
\label{Eq.: Contribution rho_R}
    \begin{split}
     \delta \mathcal{M}^{\mathrm{1-loop, \hat{\rho}_R}} 
    =& \dfrac{i}{2} \int \dfrac{d^4k}{(2 \pi)^4} \dfrac{1}{(k-p)^2} s_G c_G y_a  \dfrac{\slashed{k}+ M_T}{k^2 -M_T^2}   y_a^\dagger 
        ~,
    \end{split}
\end{equation}
which cancels due to the different signs that stem from the mixing.

Next, the contribution from the scalar exchange diagram in Fig.~\ref{Fig.: Systematic of one-loop contributions}b) is given by
\begin{equation}
	\label{Eq.: 1-Loop Correction scalar exchange}
	\begin{split}
		\delta  \mathcal{M}^{(1), \chi} &=  i \dfrac{ \lambda_{2}   v_{L} v_{R}}{2} y_a\int \dfrac{d^4 k}{(2 \pi)^4} \dfrac{ \slashed{k} +M_{T}}{ \left(k^2-M_{T}^2\right) \left((p-k)^2-m^2_{\chi_L}\right)\left((p-k)^2-m^2_{\chi_R}\right)} y^{\dagger}_{a} \\
		&= \dfrac{ \lambda_{2} v_{L} v_{R}}{32 \pi^2} y_{a}\dfrac{M_{T}}{m_{\chi_L}^2-m_{\chi_R}^2}\left[\dfrac{m_{\chi_L}^2 \log\left(\frac{M^2_{T}}{m_{\chi_L}^2}\right)}{M_{T}^2-m_{\chi_L}^2}-\dfrac{m_{\chi_R}^2 \log\left(\frac{M^2_{T}}{m_{\chi_R}^2}\right)}{M_{T}^2-m_{\chi_R}^2}\right]y^{\dagger}_{a}\, , 
	\end{split}
\end{equation}
where we evaluate all diagrams at zero external momentum from now on $p=0$. The total contribution to the upper left block in the up quark mass matrix is hence
\begin{equation}
    \delta M_u^{(1)}\equiv \delta\mathcal{M}^{(1), Z}+\delta\mathcal{M}^{(1), Z'}+\delta\mathcal{M}^{(1), \chi}~.
\end{equation}

We proceed by evaluating the diagrams in Fig.~\ref{Fig.: Systematic of one-loop contributions}~i) and ~\ref{Fig.: Systematic of one-loop contributions}~f). 
The contribution from the $X$ exchange is given by
\begin{equation}
\label{Eq.: X boson echange INT }
    \begin{split}
        \delta \mathcal{M}^{\mathrm{(1), X}}
       =&  i  \dfrac{g_X^2 \alpha^2}{4} v_\eta^2 \dfrac{y_1^\dagger}{\sqrt{2}} \int \dfrac{d^4k}{(2 \pi)^4}  \dfrac{1}{k^2} \dfrac{M_T} {k^2 -M_T^2}  ~   \dfrac{ 4}{k^2 -M_X^2} \dfrac{y_1}{\sqrt{2}}\\
        & + i  \dfrac{g_X^2 \alpha^2 }{4} \dfrac{ v_\eta^2}{M_X^2} \dfrac{y_1^\dagger}{\sqrt{2}} \int \dfrac{d^4k}{(2 \pi)^4}    \dfrac{1}{k^2}\dfrac{k^2 M_T}{k^2 -M_T^2}  ~ \left(-\dfrac{1}{k^2-M_X^2}+\dfrac{1}{k^2 -\xi_X M_X^2} \right) \dfrac{y_1}{\sqrt{2}}  ~,
    \end{split}
\end{equation}
where we use the identities  $\gamma^\mu g_{\mu \nu} \gamma^\nu =4 \cdot \identity$, $\slashed{k} \slashed{k}= k^2$ and $\gamma^\mu (k-p)_{\mu} (k-p)_{\nu} \gamma^\nu = (k-p)^2$. Inserting the  $M_X$ mass then yields
\begin{equation}
\label{Eq.: X boson echange INT 2}
    \begin{split}
        \delta \mathcal{M}^{(1), X} 
       =&  i  \dfrac{g_X^2 \alpha^2}{4} v_\eta^2 \dfrac{y_1^\dagger}{\sqrt{2}} \int \dfrac{d^4k}{(2 \pi)^4}  \dfrac{1}{k^2} \dfrac{M_T} {k^2 -M_T^2}  ~   \dfrac{ d}{k^2 -M_X^2} \dfrac{y_1}{\sqrt{2}}\\
        & +  \dfrac{i}{2}  y_1^\dagger\int \dfrac{d^4k}{(2 \pi)^4}    \dfrac{ M_T}{k^2 -M_T^2}  ~ \left(-\dfrac{1}{k^2-M_X^2}+\dfrac{1}{k^2 -\xi_X M_X^2} \right) y_1  ~.
    \end{split}
\end{equation}
The contribution from the scalar diagram with an $\eta$ exchange is given by
\begin{equation}
    \begin{split}
     \delta \mathcal{M}^{(1), \eta} 
    =& \dfrac{i}{2}    y_1^\dagger \int \dfrac{d^4k}{(2 \pi)^4}   \dfrac{ M_T}{k^2 -M_T^2} \dfrac{1}{k^2 - M_\eta^2}  y_1 
        ~.
    \end{split}
\end{equation}
From this result, we get the Goldstone boson contribution by replacing $M_\eta^2 \rightarrow \xi_X M_X^2 $ and adding an overall minus sign from the additional factor of $i$ at each vertex:
\begin{equation}
    \begin{split}
     \delta \mathcal{M}^{(1), \rho_\eta}
    =& -\dfrac{i}{2}    y_1^\dagger \int \dfrac{d^4k}{(2 \pi)^4}   \dfrac{ M_T}{k^2 -M_T^2} \dfrac{1}{k^2 - \xi_X M_X^2}  y_1 ~.
    \end{split}
\end{equation}
We note that the gauge-dependent parts from the $X$ contribution cancel exactly the Goldstone contributions. Thus, we get in total
\begin{equation}
\label{Eq.: X boson echange INT total}
    \begin{split}
        \delta \tilde{M}_u^{(1)} \equiv& ~\delta \mathcal{M}^{(1), X}+ \delta \mathcal{M}^{(1), \eta}\\
       =&~  i  \dfrac{g_X^2 \alpha^2}{4} v_\eta^2 \dfrac{y_1^\dagger}{\sqrt{2}} \int \dfrac{d^4k}{(2 \pi)^4}  \dfrac{1}{k^2} \dfrac{M_T} {k^2 -M_T^2}  ~   \dfrac{ d}{k^2 -M_X^2} \dfrac{y_1}{\sqrt{2}}\\
        & -  \dfrac{i}{2}   y_1^\dagger\int \dfrac{d^4k}{(2 \pi)^4}    \dfrac{ M_T}{k^2 -M_T^2}  ~ \dfrac{1}{k^2-M_X^2} y_1  ~\\
        & +\dfrac{i}{2}   y_1^\dagger \int \dfrac{d^4k}{(2 \pi)^4}   \dfrac{ M_T}{k^2 -M_T^2} \dfrac{1}{k^2 - M_\eta^2}  y_1 ~.
    \end{split}
\end{equation}
Carrying out the momentum integrals yields 
\begin{equation}
\label{Eq.: X boson echange INT total 2}
    \begin{split}
        \delta \tilde{M}_u^{(1)}
       =&  -i   \dfrac{g_X^2 \alpha^2}{4} v_\eta^2 \dfrac{y_1^\dagger}{2} 4 M_T \dfrac{\log(\frac{M_X^2}{M_T^2})}{M_X^2-M_T^2} y_1\\
        &+  \dfrac{1}{2}   y_1^\dagger  \dfrac{1}{16 \pi^2} M_T\left(1+ \tilde{\epsilon}- \dfrac{M_X^2 \log (\frac{M_X^2}{M_T^2})}{M_X^2-M_T^2} + \log (\mu^2/M_T^2) \right) y_1  ~\\
        & -\dfrac{1}{2}   y_1^\dagger  \dfrac{1}{16 \pi^2} M_T\left(1+ \tilde{\epsilon}- \dfrac{M_\eta^2 \log (\frac{M_\eta^2}{M_T^2})}{M_\eta^2-M_T^2} + \log (\mu^2/M_T^2) \right)  y_1 ~.
    \end{split}
\end{equation}
The expression in the first row is finite, while divergences in the second and third row cancel each other. By reordering terms, we arrive at 
\begin{equation}
\label{Eq.: X boson echange INT total 3}
    \begin{split}
        \delta \tilde{M}_u^{(1)}
       =&   \dfrac{g_X^2 v_\eta^2\alpha^2}{4 \times 32 \pi^2}  y_1^\dagger 4 M_T \dfrac{\log(\frac{M_X^2}{M_T^2})}{M_X^2-M_T^2} y_1\\
        &-    \dfrac{g_X^2 v_\eta^2 \alpha^2}{4 \times 32 \pi^2}y_1^\dagger  M_T \dfrac{ \log (\frac{M_X^2}{M_T^2})}{M_X^2-M_T^2}  y_1  ~\\
        & +  \dfrac{1}{2}  y_1^\dagger  \dfrac{1}{16 \pi^2} M_T \dfrac{M_\eta^2 \log (\frac{M_\eta^2}{M_T^2})}{M_\eta^2-M_T^2}   y_1 ~.
    \end{split}
\end{equation}
which gives the final result 
\begin{equation}
\label{Eq.: X boson echange INT total 4}
    \begin{split}
        \delta \tilde{M}_u^{(1)}
       =&   \dfrac{3}{32 \pi^2} \dfrac{g_X^2 v_\eta^2\alpha^2}{4} y_1^\dagger  M_T \dfrac{\log(\frac{M_T^2}{M_X^2})}{M_T^2-M_X^2} y_1\\
        & +   y_1^\dagger  \dfrac{1}{32 \pi^2} M_T \dfrac{M_\eta^2 \log (\frac{M_T^2}{M_\eta^2})}{M_T^2-M_\eta^2}   y_1 ~.
    \end{split}
\end{equation}
In Section \ref{SEC-03}, we summarize contributions from the diagrams Fig.~\ref{Fig.: Systematic of one-loop contributions} c) and j) by $\delta^\dagger v_R$ and those of the diagrams Fig.~\ref{Fig.: Systematic of one-loop contributions} d) and k) as $\delta v_L$. Their analytic expression can be easily deduced from the preceding results. Gauge boson couplings proportional to the identity matrix in generation space do not contribute to the masses of second and first-generation fermions and are therefore not considered in our calculation. 
\subsection*{Two-loop corrections}

The off-shell amplitude for the $W_L-W_R$ mixing in diagram \ref{Fig.: WLWR_mixing}) is given by
\begin{equation}
	\label{Eq.: W mixing amplitude}
	\begin{split}
		\Pi_{\sigma \rho }(p^2)=& -i N_c~ g_{\sigma \rho} \dfrac{g_L g_R}{2} \dfrac{v_L^2 v_R^2}{4} \sum_{\alpha, \beta = 1}^{3} \sum_{k,l = 1}^{2} \frac{[y^q_a]_{\alpha k}}{\sqrt{2}} \frac{[y^{q \dagger}_a]_{k \beta}}{\sqrt{2}} \frac{[y^q_c]_{\alpha l}}{\sqrt{2}} \frac{[y^{q \dagger }_c]_{l\beta}}{\sqrt{2}} M_{T_k} M_{B_l} \\
		\times& \int\frac{d^4k}{(2\pi)^4} \dfrac{1}{ (p+k)^2}   \frac{1}{(p+k)^2-M_{T_k}^2}  
		\dfrac{1}{ k^2} \frac{1}{k^2-M_{B_l}^2}~,
	\end{split}
\end{equation}
where we added again the superscript $q$ to the Yukawa coupling matrices in order to avoid confusion due to the numerous couplings appearing in the expression.
The previous result can now be used to calculate the two-loop radiative correction in unitary gauge (see Fig.~\ref{Fig.: Two-Loop W exchange})  that is given by
\begin{equation}
	\label{Eq.:two loop radiative correction CC}
	\begin{split}
		\delta M^{(2)}_{ij} &=  \dfrac{g_L g_R}{2}\dfrac{v_L v_R}{2} \int\dfrac{d^4 p}{(2 \pi)^4} \gamma_\mu  \dfrac{\left(g^{\mu \sigma} -p^\mu p^\sigma /M^2_{W_L}\right)}{p^2-M^2_{W_L}} \Pi_{\sigma \rho }(p^2) \\
		&\times\dfrac{\left(g^{\rho \nu} -p^\rho p^\nu /M^2_{W_R}\right)}{p^2-M^2_{W_R}} \gamma_\nu  \dfrac{1}{p^2} \frac{[y^q_c]_{ik}}{\sqrt{2}} \frac{M_{B_k}}{p^2-M_{B_k}^2}\frac{[y^{q \dagger}_c]_{kj}}{\sqrt{2}}
	\end{split}~.
\end{equation}
Together, Eqs. (\ref{Eq.: W mixing amplitude}, \ref{Eq.:two loop radiative correction CC}) yield
\begin{equation}
\label{Eq.: two-loop corrcetion}
	\begin{split}
		\delta M^{(2)}_{ij} &= N_c  \sum_{\alpha,\beta =1}^{3}\sum_{k,l,m =1}^{2} \dfrac{g^2_L g^2_R}{4}\dfrac{v^3_L v^3_R}{8} \dfrac{M_{T_k} M_{B_l}}{M_{W_L}^2 M_{W_R}^2} \frac{[y^q_a]_{\alpha k}}{\sqrt{2}} \frac{[y^{q \dagger}_a]_{k \beta}}{\sqrt{2}} \frac{[y^q_c]_{\alpha l}}{\sqrt{2}}\frac{[y^{q \dagger}_c]_{l \beta}}{\sqrt{2}} \frac{[y^q_c]_{im}}{\sqrt{2}}\frac{[y^{q \dagger}_c]_{mj}}{\sqrt{2}} M_{B_m}
		    I_{klm}~,
	\end{split}
\end{equation}

\begin{equation}
\label{Eq.: Two loop W integral equation}
\resizebox{1.0\hsize}{!}{$
		I_{klm} \equiv \int \dfrac{d^4 k}{(2 \pi)^4} \int \dfrac{d^4 p}{(2 \pi)^4}
		   \dfrac{3M_{W_L}^2 M_{W_R}^2 + (p^2-M_{W_L}^2 )(p^2-M_{W_R}^2 )}{\left((p+k)^2-M_{T_k}^2\right) \left(k^2-M_{B_l}^2\right)\left(p^2-M_{B_m}^2\right)p^2 (p+k)^2 k^2 (p^2-M_{W_L}^2 ) (p^2-M_{W_R}^2 ) }$~.}
\end{equation}
In \cite{Babu:2022ikf} it was shown that the first term in the numerator of Eq.~(\ref{Eq.: Two loop W integral equation}) is small compared to the second one and will be therefore neglected it in the following. Then the integral in equation Eq.~(\ref{Eq.: Two loop W integral equation}) evaluates to 
\begin{equation}
    \begin{split}
		I_{klm} \simeq \int \dfrac{d^4 k}{(2 \pi)^4} \int \dfrac{d^4 p}{(2 \pi)^4}
		 &  \dfrac{1}{\left((p+k)^2-M_{T_k}^2\right) \left(k^2-M_{B_l}^2\right)\left(p^2-M_{B_m}^2\right)p^2 (p+k)^2 k^2  }
	\end{split}
\end{equation}
Applying a partial fraction decomposition yields
\begin{equation}
    \begin{split}
		I_{klm} &= \dfrac{1}{M_{B_l}^2 M_{T_k}^2 M_{B_m}^2}\int \dfrac{d^4 k}{(2 \pi)^4} \left[\dfrac{1}{\left(k^2-M_{B_l}^2\right)}-\dfrac{1}{k^2}\right] \\
  &\int \dfrac{d^4 p}{(2 \pi)^4}
		  \left[ \dfrac{1}{\left((p+k)^2-M_{T_k}^2\right)} -\dfrac{1}{(p+k)^2}\right] \left[ \dfrac{1}{\left(p^2-M_{B_m}^2\right)}- \dfrac{1}{p^2}\right] ~.
	\end{split}
\end{equation}
Following the calculation method along \cite{vanderBij:1983bw,Babu:2022ikf} and multiplying out all terms one obtains
\begin{math}
\begin{aligned}
\label{Eq.: Partial fraction Decomposition}
		I_{klm} &= \dfrac{1}{M_{B_l}^2 M_{T_k}^2 M_{B_m}^2}\int \dfrac{d^4 k}{(2 \pi)^4} \int \dfrac{d^4 p}{(2 \pi)^4} \left[
		\dfrac{1}{\left( k^2-M_{B_l}^2\right)\left((p+k)^2-M_{T_k}^2\right)\left(p^2-M_{B_m}^2\right)} \right. \\  
		& \left. -\dfrac{1}{\left(k^2-M_{B_l}^2\right)\left((p+k)^2-M_{T_k}^2\right)p^2}
		-\dfrac{1}{\left(k^2-M_{B_l}^2\right) \left(p+k\right)^2\left(p^2-M_{B_m}^2\right)}
		+\dfrac{1}{\left(k^2-M_{B_l}^2\right)\left(p+k\right)^2 p^2} \right. \\
		& \left. -\dfrac{1}{k^2\left((p+k)^2-M_{T_k}^2\right)\left(p^2-M_{B_m}^2\right)}
		+\dfrac{1}{k^2\left((p+k)^2-M_{T_k}^2\right)p^2} \right.\\
		& \left. +\dfrac{1}{k^2 \left(p+k\right)^2 \left(p^2-M_{B_m}^2\right) }
		-\dfrac{1}{k^2 \left(p+k\right)^2 p^2} \right]~.
\end{aligned}
\end{math}
For better readability we define the notation
\begin{align*}
        &\left(M_{1_1} M_{1_2}....M_{1_{n_1}}| M_{2_1} ....M_{2_{n_2}}| M_{3_1} ....M_{3_{n_3}} \right) \\
        &= \int d^n k \int d^n p \prod_{i=1}^{n_1}\prod_{j=1}^{n_2}\prod_{i=l}^{n_3} \dfrac{1}{\left( k^2-M_{1_i}^2\right)} \dfrac{1}{\left( p^2-M_{2_j}^2\right)}
        \dfrac{1}{\left( (k+p)^2-M_{3_l}^2\right)}
\end{align*}
such that the whole expression can be rewritten in short form as 
\begin{align*}
    I_{klm} = \dfrac{1}{M_{B_l}^2 M_{T_k}^2 M_{B_m}^2}\dfrac{1}{(2\pi)^8} &\left[ (M_{B_l}|M_{B_m}|M_{T_k})- (M_{B_l}|0|M_{T_k})-(M_{B_l}|M_{B_m}|0)+(M_{B_l}|0|0) \right. \\ 
    &\left. -(0|M_{B_m}|M_{T_k})+(0|0|M_{T_k}) +(0|M_{B_m}|0)-(0|0|0)
    \right]~.
\end{align*}
Now, using the identity
\begin{align*}
    \left(M_0|M_1|M_2 \right) = \dfrac{1}{3-n}\left[ 
    M_0^2\left(M_0 M_0| M_1|M_2 \right) +M_1^2\left(M_1 M_1| M_0|M_2 \right) +M_2^2\left(M_2 M_2| M_0|M_1 \right) 
    \right]~,
\end{align*}
with $n=4$ being the dimension of the integral, we find
\begin{align*}
    I_{klm} &= \dfrac{1}{M_{B_l}^2 M_{T_k}^2 M_{B_m}^2}\dfrac{1}{(2\pi)^8} \dfrac{1}{3-n} \left[
    M_{B_l}^2 (M_{B_l} M_{B_l}|M_{B_m}|M_{T_k}) \right. \\ 
    & \left. + 
     M_{B_m}^2 (M_{B_m} M_{B_m}|M_{B_l}|M_{T_k})+
      M_{T_k}^2 (M_{T_k} M_{T_k}|M_{B_l}|M_{B_m}) \right. \\
      & \left. - M_{B_l}^2 (M_{B_l} M_{B_l}|0|M_{T_k})
      - M_{T_k}^2 (M_{T_k} M_{T_k}|M_{B_l}|0)
      - M_{B_l}^2 (M_{B_l} M_{B_l}|M_{B_m}|0) \right. \\
      & \left. - M_{B_m}^2 (M_{B_m} M_{B_m}|M_{B_l}|0)
      + M_{B_l}^2 (M_{B_l} M_{B_l}|0|0)
      - M_{B_m}^2 (M_{B_m} M_{B_m}|0|M_{T_k}) \right. \\
      & \left. -M_{T_k}^2 (M_{T_k} M_{T_k}|0|M_{B_m})
      +M_{T_k}^2 (M_{T_k} M_{T_k}|0|0)
      +M_{B_m}^2 (M_{B_m} M_{B_m}|0|0) \right]~.
\end{align*}
Finally, we can  use \cite{vanderBij:1983bw}
\begin{align*}
    (M M |M_1| M_2) = &\pi^4 \left[ 
    \dfrac{-2}{\epsilon^2} + \dfrac{1}{\epsilon} \left( 1-2\gamma_E -2 \log(\pi M^2)\right)
    \right]\\
    +&\pi^4 \left[-\dfrac{1}{2} -\dfrac{1}{12}\pi^2 + \gamma_E - \gamma_E^2 + (1-2\gamma_E) \log(\pi M^2) -\log(\pi M^2)^2 -f(a,b) \right]  \\
    &+  \mathcal{O}(\epsilon^2)~,
\end{align*}
where $a = M_1^2/M^2$, $b = M_2^2/M^2$ and the function $f$ is defined as
\begin{align*}
    f(a,b) &= -\dfrac{1}{2} \log(a) \log(b) - \dfrac{1}{2} \left( \dfrac{a+b-1}{\sqrt{\Delta}}\right) \left[
    \mathrm{Li}_2 \left(\dfrac{-x_2}{y_1}\right) + \mathrm{Li}_2 \left(\dfrac{-y_2}{x_1}\right)  -\mathrm{Li}_2 \left(\dfrac{-x_1}{y_2}\right) \right. \\
    & \left. -\mathrm{Li}_2 \left(\dfrac{-y_1}{x_2}\right) +\mathrm{Li}_2 \left(\dfrac{b-a}{x_2}\right) + \mathrm{Li}_2 \left(\dfrac{a-b}{y_2}\right)-\mathrm{Li}_2 \left(\dfrac{b-a}{x_1}\right)  -\mathrm{Li}_2 \left(\dfrac{a-b}{y_1}\right)  
    \right]~,
\end{align*}
with $\mathrm{Li}_2(x)$ the dilogarithm function, $\Delta = 1-2(a+b) +(a-b)^2$ and 
\begin{align*}
    x_1 &= \dfrac{1}{2} \left( 1+b-a +\sqrt{\Delta}\right)~, \\
     x_2 &= \dfrac{1}{2} \left( 1+b-a -\sqrt{\Delta}\right) ~,\\
      y_1 &= \dfrac{1}{2} \left( 1+a -b+\sqrt{\Delta}\right) ~,\\
       y_2 &= \dfrac{1}{2} \left( 1+a-b- \sqrt{\Delta}\right) ~.
\end{align*}
This allows us to rewrite the integral in the form of
\begin{align*}
     I_{klm} &= \dfrac{1}{ M_{T_k}^2 M_{B_l}^2} \dfrac{1}{(2\pi)^8} \dfrac{1}{3-n} \pi^4
     \left[ -f(r_1,r_2) + f(r_1,0) + f(0, r_2) -f(0,0) \right. \\
     &+ \left. r_1 \left( 
     -f(1/r_1, r_2/r_1) +f(0,r_2/r_1) + f(1/r_1,0) -f(0,0)
     \right)\right. \\
     &+ \left. r_2 \left(
     -f(r_1/r_2, 1/r_1) +f(r_1/r_2, 0) + f(0,1/r_2) -f(0,0)
     \right)
     \right]~,
\end{align*}
where $r_1 = M_{B_l}^2/ M_{B_m}^2 $ and $r_2 = M_{T_k}^2/ M_{B_m}^2$.
Note that the defined function  
 $f$ has the properties $f(a,b) = f(b,a)$, $f(0,0) = \pi^2/6$ and $f(a,0) = \mathrm{Li}_2 \left(1-a \right)$. This yields the final result
\begin{align*}
    I_{klm} &= -\dfrac{1}{(16\pi^2)^2} \dfrac{1}{ M_{T_k}^2 M_{B_l}^2} \left[
    -\dfrac{\pi^2}{6}\left( 1+r_1 +r_2\right) - f(r_1,r_2) + \mathrm{Li}_2\left(1-r_1\right)  + \mathrm{Li}_2\left(1-r_2\right) \right. \\
    & \left. +r_1 \left(-f\left(\dfrac{1}{r_1}, \dfrac{r_2}{r_1}\right) + \mathrm{Li}_2\left(1-\dfrac{r_2}{r_1}\right)+ \mathrm{Li}_2\left(1-\dfrac{1}{r_1}\right) \right) \right. \\
    & \left. +r_2 \left(-f\left(\dfrac{r_1}{r_2}, \dfrac{1}{r_2}\right) + \mathrm{Li}_2\left(1-\dfrac{r_1}{r_2}\right)+ \mathrm{Li}_2\left(1-\dfrac{1}{r_2} \right) \right)
    \right]~.
\end{align*}
Together with Eq. (\ref{Eq.: two-loop corrcetion}) this gives

\begin{align}
\label{Eq.: Two-loop correction Final Result}
    	\delta M^{(2)}_{ij} = -\dfrac{N_c}{(16 \pi^2)^2} &\sum_{\alpha, \beta =1}^{3} \sum_{k,l,m =1}^{2} \dfrac{g^2_L g^2_R}{4}\dfrac{v^3_L v^3_R}{8} \dfrac{M_{B_m} }{M_{T_k} M_{B_l} M_{W_L}^2 M_{W_R}^2} \frac{[y^q_a]_{\alpha k}}{\sqrt{2}} \frac{[y^{q \dagger}_a]_{k \beta}}{\sqrt{2}} \frac{[y^q_c]_{\alpha l}}{\sqrt{2}}\frac{[y^{q \dagger}_c]_{l \beta}}{\sqrt{2}} \frac{[y^q_c]_{im}}{\sqrt{2}}\frac{[y^{q \dagger}_c]_{mj}}{\sqrt{2}} \nonumber \\
		  &\left[
    -\dfrac{\pi^2}{6}\left( 1+r_1 +r_2\right) - f(r_1,r_2) + \mathrm{Li}_2\left(1-r_1\right)  + \mathrm{Li}_2\left(1-r_2\right) \right. 
 \nonumber \\
    & \left. +r_1 \left(-f\left(\dfrac{1}{r_1}, \dfrac{r_2}{r_1}\right) + \mathrm{Li}_2\left(1-\dfrac{r_2}{r_1}\right)+ \mathrm{Li}_2\left(1-\dfrac{1}{r_1}\right) \right) \right. \nonumber \\
    & \left. +r_2 \left(-f\left(\dfrac{r_1}{r_2}, \dfrac{1}{r_2}\right) + \mathrm{Li}_2\left(1-\dfrac{r_1}{r_2}\right)+ \mathrm{Li}_2\left(1-\dfrac{1}{r_2} \right) \right)
    \right]~.
\end{align}


\clearpage
\bibliographystyle{utphys}
\bibliography{reference}

\providecommand{\href}[2]{#2}\begingroup\raggedright\begin{thebibliography}{10}

\bibitem{Froggatt:1978nt}
C.~D. Froggatt and H.~B. Nielsen, ``{Hierarchy of Quark Masses, Cabibbo Angles and CP Violation},'' \href{http://dx.doi.org/10.1016/0550-3213(79)90316-X}{{\em Nucl. Phys. B} {\bfseries 147} (1979) 277--298}.

\bibitem{Arkani-Hamed:1999ylh}
N.~Arkani-Hamed and M.~Schmaltz, ``{Hierarchies without symmetries from extra dimensions},'' \href{http://dx.doi.org/10.1103/PhysRevD.61.033005}{{\em Phys. Rev. D} {\bfseries 61} (2000) 033005}, \href{http://arxiv.org/abs/hep-ph/9903417}{{\ttfamily arXiv:hep-ph/9903417}}.

\bibitem{Babu:2009fd}
K.~S. Babu, \href{http://dx.doi.org/10.1142/9789812838360_0002}{``{TASI Lectures on Flavor Physics},''} in {\em {Theoretical Advanced Study Institute in Elementary Particle Physics}: {The Dawn of the LHC Era}}, pp.~49--123.
\newblock 2010.
\newblock \href{http://arxiv.org/abs/0910.2948}{{\ttfamily arXiv:0910.2948 [hep-ph]}}.

\bibitem{King:2013eh}
S.~F. King and C.~Luhn, ``{Neutrino Mass and Mixing with Discrete Symmetry},'' \href{http://dx.doi.org/10.1088/0034-4885/76/5/056201}{{\em Rept. Prog. Phys.} {\bfseries 76} (2013) 056201}, \href{http://arxiv.org/abs/1301.1340}{{\ttfamily arXiv:1301.1340 [hep-ph]}}.

\bibitem{Balakrishna:1987qd}
B.~S. Balakrishna, ``{Fermion Mass Hierarchy From Radiative Corrections},'' \href{http://dx.doi.org/10.1103/PhysRevLett.60.1602}{{\em Phys. Rev. Lett.} {\bfseries 60} (1988) 1602}.

\bibitem{Balakrishna:1988ks}
B.~S. Balakrishna, A.~L. Kagan, and R.~N. Mohapatra, ``{Quark Mixings and Mass Hierarchy From Radiative Corrections},'' \href{http://dx.doi.org/10.1016/0370-2693(88)91676-0}{{\em Phys. Lett. B} {\bfseries 205} (1988) 345--352}.

\bibitem{Balakrishna:1988bn}
B.~S. Balakrishna and R.~N. Mohapatra, ``{Radiative Fermion Masses From New Physics at Tev Scale},'' \href{http://dx.doi.org/10.1016/0370-2693(89)91129-5}{{\em Phys. Lett. B} {\bfseries 216} (1989) 349--352}.

\bibitem{Babu:1989tv}
K.~S. Babu, B.~S. Balakrishna, and R.~N. Mohapatra, ``{Supersymmetric Model for Fermion Mass Hierarchy},'' \href{http://dx.doi.org/10.1016/0370-2693(90)91433-C}{{\em Phys. Lett. B} {\bfseries 237} (1990) 221--228}.

\bibitem{Babu:1988fn}
K.~S. Babu and X.-G. He, ``{Fermion mass hierarchy and the strong CP problem},'' \href{http://dx.doi.org/10.1016/0370-2693(89)90401-2}{{\em Phys. Lett. B} {\bfseries 219} (1989) 342--346}.

\bibitem{Rattazzi:1990wu}
R.~Rattazzi, ``{Radiative quark masses constrained by the gauge group only},'' \href{http://dx.doi.org/10.1007/BF01562331}{{\em Z. Phys. C} {\bfseries 52} (1991) 575--582}.

\bibitem{Dobrescu:2008sz}
B.~A. Dobrescu and P.~J. Fox, ``{Quark and lepton masses from top loops},'' \href{http://dx.doi.org/10.1088/1126-6708/2008/08/100}{{\em JHEP} {\bfseries 08} (2008) 100}, \href{http://arxiv.org/abs/0805.0822}{{\ttfamily arXiv:0805.0822 [hep-ph]}}.

\bibitem{Weinberg:2020zba}
S.~Weinberg, ``{Models of Lepton and Quark Masses},'' \href{http://dx.doi.org/10.1103/PhysRevD.101.035020}{{\em Phys. Rev. D} {\bfseries 101} no.~3, (2020) 035020}, \href{http://arxiv.org/abs/2001.06582}{{\ttfamily arXiv:2001.06582 [hep-th]}}.

\bibitem{Mohanta:2022seo}
G.~Mohanta and K.~M. Patel, ``{Radiatively generated fermion mass hierarchy from flavour non-universal gauge symmetries},'' \href{http://arxiv.org/abs/2207.10407}{{\ttfamily arXiv:2207.10407 [hep-ph]}}.

\bibitem{Mohanta:2023soi}
G.~Mohanta and K.~M. Patel, ``{Gauged SU(3)$_{F}$ and loop induced quark and lepton masses},'' \href{http://dx.doi.org/10.1007/JHEP10(2023)128}{{\em JHEP} {\bfseries 10} (2023) 128}, \href{http://arxiv.org/abs/2308.05642}{{\ttfamily arXiv:2308.05642 [hep-ph]}}.

\bibitem{Graham:2009gr}
P.~W. Graham and S.~Rajendran, ``{A Domino Theory of Flavor},'' \href{http://dx.doi.org/10.1103/PhysRevD.81.033002}{{\em Phys. Rev. D} {\bfseries 81} (2010) 033002}, \href{http://arxiv.org/abs/0906.4657}{{\ttfamily arXiv:0906.4657 [hep-ph]}}.

\bibitem{Jana:2021tlx}
S.~Jana, S.~Klett, and M.~Lindner, ``{Flavor seesaw mechanism},'' \href{http://dx.doi.org/10.1103/PhysRevD.105.115015}{{\em Phys. Rev. D} {\bfseries 105} no.~11, (2022) 115015}, \href{http://arxiv.org/abs/2112.09155}{{\ttfamily arXiv:2112.09155 [hep-ph]}}.

\bibitem{Bonilla:2023wok}
C.~Bonilla, A.~E. Carcamo~Hernandez, S.~Kovalenko, H.~Lee, R.~Pasechnik, and I.~Schmidt, ``{Fermion mass hierarchy in an extended left-right symmetric model},'' \href{http://dx.doi.org/10.1007/JHEP12(2023)075}{{\em JHEP} {\bfseries 12} (2023) 075}, \href{http://arxiv.org/abs/2305.11967}{{\ttfamily arXiv:2305.11967 [hep-ph]}}.

\bibitem{Mohanta:2024wcr}
G.~Mohanta and K.~M. Patel, ``{Loop-induced masses for the first two generations with optimum flavour violation},'' \href{http://arxiv.org/abs/2406.19179}{{\ttfamily arXiv:2406.19179 [hep-ph]}}.

\bibitem{Berezhiani:1983hm}
Z.~G. Berezhiani, ``{The Weak Mixing Angles in Gauge Models with Horizontal Symmetry: A New Approach to Quark and Lepton Masses},'' \href{http://dx.doi.org/10.1016/0370-2693(83)90737-2}{{\em Phys. Lett. B} {\bfseries 129} (1983) 99--102}.

\bibitem{Chang:1986bp}
D.~Chang and R.~N. Mohapatra, ``{Small and Calculable Dirac Neutrino Mass},'' \href{http://dx.doi.org/10.1103/PhysRevLett.58.1600}{{\em Phys. Rev. Lett.} {\bfseries 58} (1987) 1600}.

\bibitem{Davidson:1987mh}
A.~Davidson and K.~C. Wali, ``{Universal Seesaw Mechanism?},'' \href{http://dx.doi.org/10.1103/PhysRevLett.59.393}{{\em Phys. Rev. Lett.} {\bfseries 59} (1987) 393}.

\bibitem{Rajpoot:1987fca}
S.~Rajpoot, ``{See-saw masses for quarks and leptons in an ambidextrous electroweak interaction model},'' \href{http://dx.doi.org/10.1142/S0217732387000422}{{\em Mod. Phys. Lett. A} {\bfseries 2} no.~5, (1987) 307--315}. [Erratum: Mod.Phys.Lett.A 2, 541 (1987)].

\bibitem{Babu:1988mw}
K.~S. Babu and R.~N. Mohapatra, ``{{CP} Violation in Seesaw Models of Quark Masses},'' \href{http://dx.doi.org/10.1103/PhysRevLett.62.1079}{{\em Phys. Rev. Lett.} {\bfseries 62} (1989) 1079}.

\bibitem{Babu:1989rb}
K.~S. Babu and R.~N. Mohapatra, ``{A Solution to the Strong {CP} Problem Without an Axion},'' \href{http://dx.doi.org/10.1103/PhysRevD.41.1286}{{\em Phys. Rev. D} {\bfseries 41} (1990) 1286}.

\bibitem{Mohapatra:2014qva}
R.~N. Mohapatra and Y.~Zhang, ``{TeV Scale Universal Seesaw, Vacuum Stability and Heavy Higgs},'' \href{http://dx.doi.org/10.1007/JHEP06(2014)072}{{\em JHEP} {\bfseries 06} (2014) 072}, \href{http://arxiv.org/abs/1401.6701}{{\ttfamily arXiv:1401.6701 [hep-ph]}}.

\bibitem{Patra:2017gak}
A.~Patra and S.~K. Rai, ``{Lepton-specific universal seesaw model with left-right symmetry},'' \href{http://dx.doi.org/10.1103/PhysRevD.98.015033}{{\em Phys. Rev. D} {\bfseries 98} no.~1, (2018) 015033}, \href{http://arxiv.org/abs/1711.00627}{{\ttfamily arXiv:1711.00627 [hep-ph]}}.

\bibitem{Chen:2022wvk}
S.-P. Chen and P.-H. Gu, ``{U(1)Y' universal seesaw},'' \href{http://dx.doi.org/10.1016/j.nuclphysb.2022.116057}{{\em Nucl. Phys. B} {\bfseries 986} (2023) 116057}, \href{http://arxiv.org/abs/2211.01906}{{\ttfamily arXiv:2211.01906 [hep-ph]}}.

\bibitem{Dcruz:2022rjg}
R.~Dcruz and K.~S. Babu, ``{Resolving W boson mass shift and CKM unitarity violation in left-right symmetric models with a universal seesaw mechanism},'' \href{http://dx.doi.org/10.1103/PhysRevD.108.095011}{{\em Phys. Rev. D} {\bfseries 108} no.~9, (2023) 095011}, \href{http://arxiv.org/abs/2212.09697}{{\ttfamily arXiv:2212.09697 [hep-ph]}}.

\bibitem{Morozumi:2024mit}
T.~Morozumi and A.~H. Panuluh, ``{The third family quark mass hierarchy and FCNC in the universal seesaw model},'' \href{http://arxiv.org/abs/2407.00732}{{\ttfamily arXiv:2407.00732 [hep-ph]}}.

\bibitem{Pati:1974yy}
J.~C. Pati and A.~Salam, ``{Lepton Number as the Fourth Color},'' \href{http://dx.doi.org/10.1103/PhysRevD.10.275}{{\em Phys. Rev. D} {\bfseries 10} (1974) 275--289}. [Erratum: Phys.Rev.D 11, 703--703 (1975)].

\bibitem{Mohapatra:1974hk}
R.~N. Mohapatra and J.~C. Pati, ``{Left-Right Gauge Symmetry and an Isoconjugate Model of CP Violation},'' \href{http://dx.doi.org/10.1103/PhysRevD.11.566}{{\em Phys. Rev. D} {\bfseries 11} (1975) 566--571}.

\bibitem{Mohapatra:1974gc}
R.~N. Mohapatra and J.~C. Pati, ``{A Natural Left-Right Symmetry},'' \href{http://dx.doi.org/10.1103/PhysRevD.11.2558}{{\em Phys. Rev. D} {\bfseries 11} (1975) 2558}.

\bibitem{Senjanovic:1975rk}
G.~Senjanovic and R.~N. Mohapatra, ``{Exact Left-Right Symmetry and Spontaneous Violation of Parity},'' \href{http://dx.doi.org/10.1103/PhysRevD.12.1502}{{\em Phys. Rev. D} {\bfseries 12} (1975) 1502}.

\bibitem{Babu:2022ikf}
K.~S. Babu, X.-G. He, M.~Su, and A.~Thapa, ``{Naturally Light Dirac and Pseudo-Dirac Neutrinos from Left-Right Symmetry},'' \href{http://arxiv.org/abs/2205.09127}{{\ttfamily arXiv:2205.09127 [hep-ph]}}.

\bibitem{Babu:2018vrl}
K.~S. Babu, B.~Dutta, and R.~N. Mohapatra, ``{A theory of R(D$^{*}$, D) anomaly with right-handed currents},'' \href{http://dx.doi.org/10.1007/JHEP01(2019)168}{{\em JHEP} {\bfseries 01} (2019) 168}, \href{http://arxiv.org/abs/1811.04496}{{\ttfamily arXiv:1811.04496 [hep-ph]}}.

\bibitem{Dev:2012sg}
P.~S.~B. Dev and A.~Pilaftsis, ``{Minimal Radiative Neutrino Mass Mechanism for Inverse Seesaw Models},'' \href{http://dx.doi.org/10.1103/PhysRevD.86.113001}{{\em Phys. Rev. D} {\bfseries 86} (2012) 113001}, \href{http://arxiv.org/abs/1209.4051}{{\ttfamily arXiv:1209.4051 [hep-ph]}}.

\bibitem{Zyla:2020zbs}
{\bfseries Particle Data Group} Collaboration, P.~A. Zyla {\em et~al.}, ``{Review of Particle Physics},'' \href{http://dx.doi.org/10.1093/ptep/ptaa104}{{\em PTEP} {\bfseries 2020} no.~8, (2020) 083C01}.

\bibitem{Esteban:2020cvm}
I.~Esteban, M.~C. Gonzalez-Garcia, M.~Maltoni, T.~Schwetz, and A.~Zhou, ``{The fate of hints: updated global analysis of three-flavor neutrino oscillations},'' \href{http://dx.doi.org/10.1007/JHEP09(2020)178}{{\em JHEP} {\bfseries 09} (2020) 178}, \href{http://arxiv.org/abs/2007.14792}{{\ttfamily arXiv:2007.14792 [hep-ph]}}.

\bibitem{Patel:2016fam}
H.~H. Patel, ``{Package-X 2.0: A Mathematica package for the analytic calculation of one-loop integrals},'' \href{http://dx.doi.org/10.1016/j.cpc.2017.04.015}{{\em Comput. Phys. Commun.} {\bfseries 218} (2017) 66--70}, \href{http://arxiv.org/abs/1612.00009}{{\ttfamily arXiv:1612.00009 [hep-ph]}}.

\bibitem{vanderBij:1983bw}
J.~van~der Bij and M.~J.~G. Veltman, ``{Two Loop Large Higgs Mass Correction to the rho Parameter},'' \href{http://dx.doi.org/10.1016/0550-3213(84)90284-0}{{\em Nucl. Phys. B} {\bfseries 231} (1984) 205--234}.

\end{thebibliography}\endgroup
\end{document}